\documentclass[universe,article,accept,pdftex,moreauthors]{Definitions/mdpi} 

\firstpage{1} 
\pubvolume{1}
\issuenum{1}
\articlenumber{0}
\pubyear{2025}
\copyrightyear{2025}
\externaleditor{Firstname Lastname}
\datereceived{ } 
\daterevised{ } 
\dateaccepted{ } 
\datepublished{ } 
\hreflink{https://doi.org/} 

\usepackage{color}
\usepackage[normalem]{ulem}
\usepackage{changes}

\Title{Parameter Estimation Precision with Geocentric Gravitational Wave Interferometers: Monochromatic Signals}

\TitleCitation{Parameter Estimation Precision with Geocentric Gravitational Wave Interferometers: Monochromatic Signals}

\Author{Manoel Felipe Sousa~$^{1,2}$\orcidA{}, Tabata Aira Ferreira~$^{3}$\orcidB{} and  Massimo Tinto~$^{4,}$*\orcidC{}}

\AuthorNames{Manoel Felipe Sousa, Tabata Aira Ferreira and  Massimo Tinto}

\isAPAStyle{%
       \AuthorCitation{Sousa, M.F., Ferreira, T.A., \& Tinto, M.}
         }{%
        \isChicagoStyle{%
        \AuthorCitation{Sousa, Manoel Felipe, Tabata Aira Ferreira, and Massimo Tinto.}
        }{
        \AuthorCitation{Sousa, M.F.; Ferreira, T.A.; Tinto, M.}
        }
}

\address{%
$^{1}$ \quad Programa de Pós-Graduação em Física e Astronomia, Universidade Tecnológica Federal do Paraná, \mbox{Curitiba 80230-901, PR, Brazil;} manoelsousa@utfpr.edu.br 
\\
$^{2}$ \quad Departamento de Engenharias e Exatas, Universidade Federal do Paraná, \mbox{Palotina 85950-000, PR, Brazil}\\
$^{3}$ \quad Department of Physics and Astronomy, Louisiana State University,
 \mbox{Baton Rouge, LA 70803, USA;} tferreira@lsu.edu \\
$^{4}$ \quad Divis\~{a}o de Astrof\'{i}sica, Instituto Nacional de Pesquisas Espaciais, \mbox{S. J. Campos 12227-010, SP, Brazil}}

\corres{Correspondence: massimo.tinto@gmail.com}

\abstract{We present a Fisher information matrix study of the
  parameter estimation precision achievable by a class of future
  space-based, ``mid-band'', gravitational wave interferometers
  observing monochromatic signals. The mid-band is the frequency
  region between that accessible by the Laser Interferometer Space
  Antenna (LISA) and ground-based interferometers. We analyze
  monochromatic signals observed by the TianQin mission, gLISA (a~LISA-like interferometer in a geosynchronous orbit) and a descoped
  gLISA mission, gLISA$_d$, characterized by an acceleration noise
  level that is three orders of magnitude worse than that of gLISA. We
  find that all three missions achieve their best angular source
  reconstruction precision in the higher part of their accessible
  frequency band, with an error box better than $10^{-10}$ sr in the
  frequency band [$10^{-1}, 10$] Hz when observing a monochromatic
  gravitational wave signal of amplitude $h_0 = 10^{-21}$ that is incoming
  from a given direction. 
  In terms of their reconstructed frequencies
  and amplitudes, TianQin achieves its best precision values in both
  quantities in the frequency band [$10^{-2}, 4 \times 10^{-1}$] Hz,
  with a frequency precision $\sigma_{f_{gw}} = 2 \times 10^{-11}$ Hz
  and an amplitude precision $\sigma_{h_0} = 2 \times 10^{-24}$. gLISA
  matches these precisions in a frequency band slightly higher than
  that of TianQin, [$3 \times 10^{-2}, 1$] Hz, as a consequence of its
  smaller arm length. gLISA$_d$, on the other hand, matches the
  performance of gLISA only over the narrower frequency region,
  [$7 \times 10^{-1}, 1$] Hz, as a consequence of its higher
  acceleration noise at lower frequencies. The angular, frequency, and
  amplitude precisions as functions of the source sky location are
  then derived by assuming an average signal-to-noise ratio of 10 at
  a selected number of gravitational wave frequencies covering the
  operational bandwidth of TianQin and gLISA. Similar precision
  functions are then derived for gLISA$_d$ by using the amplitudes
  resulting in the gLISA average SNR being equal to 10 at the selected
  frequencies. We find that, for {\underline {any}} given source
  location, all three missions display a marked precision improvement
  in the three reconstructed parameters at higher {gravitational wave} frequencies.  }

\keyword{gravitational waves; space-based interferometry; data
  analysis; monochromatic gravitational wave signals}

\begin{document}
\section{Introduction}
\label{SecI}

The first direct observation of a Gravitational Wave (GW) signal was
announced by the Laser Interferometer Gravitational-Wave Observatory
(LIGO) project~\cite{LIGO} on 11 February 2016~\cite{GW150914}. This
event, named GW150914, represents one of the most important
achievements in experimental physics today. Two interferometers, located in Livingston (Louisiana) and Hanford (Washington),
  simultaneously measured and recorded strain data, providing
  researchers with a remarkable level of confidence in the
  detection. This allowed them to conclusively identify the source of
  the observed {GW} signal as a merging binary system of
  black holes, with~component masses of $M_1 = 36^{+5}_{-4} \ M_\odot$
  and $M_2 = 29^{+4}_{-4} \ M_\odot$. The~event was detected at a
  luminosity distance of $410^{+160}_{-180}$~Mpc, corresponding to a
  redshift of $z = 0.09^{+0.03}_{-0.04}$, with~uncertainties reported
  at the $90 \%$ confidence~level.

The direct observation of this event marks the beginning of GW
astronomy~\cite{Thorne1987}, a~historic moment comparable in
magnitude to the early astronomical observations made in the year 1610
by Galileo Galilei~\cite{Galileo}. Quite like Galileo then, we
have just started to explore the capabilities of our new
observational tools, which promises to reveal secrets of the universe
inaccessible by any other~means.

Since the first detection announcement in 2016, several other GW
signals have been observed by the LIGO--Virgo--KAGRA (LVK) collaboration
~\cite{LIGO,VIRGO}. Ground-based detectors that are widely separated on
Earth and operate in coincidence can discriminate a GW signal from
random noise and provide enough information for reconstructing the
source's sky location, luminosity distance, mass(es), dynamic time
scale, and~other observables~\cite{SchutzTinto,GT}.

Space-based interferometers, on~the other hand, have enough data
redundancy to validate their measurements and uniquely reconstruct an
observed signal with their six links along their three arms~\cite{TD2020,LISA2017}. Missions such as LISA~\endnote{The LISA
  mission went through several design assessments over its development
  cycle, each resulting in a different mission sensitivity. Here, we
  will adopt the latest LISA mission concept, characterized by an arm
  length of $2.5$~Mkm, an~acceleration noise of
  $3.0 \times 10^{-15} \ \rm m/s^2 \ (Hz)^{-1/2}$, and a high-frequency
  noise of $12 \ \rm pm / \sqrt{Hz}$.} or the Chinese mission TaiJi~\cite{TaiJi}, with~their million-kilometers-long optical links, will
be able to estimate the phase noise levels and its statistical
properties over the observational frequency bands they operate
within. By~relying on a Time-Delay Interferometric (TDI) measurement~\cite{TD2020} that is insensitive to GWs~\cite{TAE01}, space-based
interferometers will assess their in-flight noise characteristics in
the lower part of the band, that is, at~frequencies smaller than the
inverse of the round-trip light time. Instead, at~higher frequencies
where they can synthesize three independent interferometric
measurements, they will be able to perform a data consistency test
based on the null stream technique~\cite{GT, TL, SST2009}, that is, a~non-linear parametric combination of the TDI measurements that
achieves a pronounced minimum at a unique point in the search
parameter space when a signal is present.  In~addition, by~taking
advantage of the Doppler and amplitude modulations introduced by the
motion of the array around the Sun on long-lived GW signals,
space-based interferometers will measure the values of the parameters
associated with the GW source of the observed signal~\cite{LISA2017}.

Although a space-based array such as LISA and TaiJi can synthesize the
equivalent of four interferometric TDI combinations (the Sagnac TDI
combinations $(\alpha, \beta, \gamma, \zeta)$, for~example)
\cite{TD2020}, their best sensitivity levels are achieved only over a
relatively narrow region of the mHz frequency band.  At~frequencies
lower than the inverse of the round-trip light time, the~sensitivity
of a space-based GW interferometer is determined by the level of
residual acceleration noise associated with the nearly free-floating
proof masses of the onboard gravitational reference sensor and the
size of the arm length. In~this region of the accessible frequency
band, the~magnitude of a GW signal in the interferometric data scales,
in fact, linearly with arm length. Instead, at~frequencies higher than
the inverse of the round-trip light time, the~sensitivity is primarily
determined by the photon count statistics at the photodetectors
~\cite{TAAA}.  The~sensitivity in this part of the accessible frequency
degrades linearly with the arm length because the shot-noise is
inversely proportional to the square root of the received optical
power and the GW signal no longer scales with the arm length. From~the
above considerations, we may conclude that, for~a defined
configuration of the on-board science instrumentation, the~best
sensitivity level and the corresponding bandwidth over which it is
achieved are uniquely determined by the size of the~array.

The frequency range over~which the best sensitivity level of a
space-based interferometer is achieved is particularly important when
detecting signals that increase in frequency over time, such as those
produced by merging binary black hole systems.  Astrophysical models
theoretically predict~\cite{Sesana} a vast population of coalescing
binary systems, with~masses similar to those involved in
GW150914. They will generate~{GWs} with characteristic
amplitudes detectable by both LISA and TaiJi within a frequency range
spanning approximately $1.5 \times 10^{-2}$ Hz to $7.6 \times 10^{-2}$
Hz.  The~lower frequency limit corresponds to the assumption of
observing a GW150914-like signal for a period of five years
(approximately equal to its coalescing time). The~upper limit instead
corresponds to the value at which the signal's amplitude equals the
interferometer's sensitivity, in~this case that of LISA. Although~one
could in principle increase the size of the optical telescopes and
rely on more powerful lasers so as to increase the upper frequency
cut-off to enlarge the observational bandwidth, in~practice, pointing
accuracy and stability requirements together with the finiteness of
the on-board available power would result in a negligible~gain.

A natural way to broaden the millihertz band, so as to fill the
frequency gap between the region accessible by LISA and TaiJi and that
by ground interferometers, is to fly additional interferometers of
smaller arm length. An~interferometer such as that of the Chinese TianQin
mission~\cite{TianQin}, or~the geosynchronous Laser Interferometer
Space Antenna (gLISA)~\cite{TAAA,TBDT,TAKAA}, could naturally
accomplish this scientific~objective.

In this article, we present an analysis of the precision achievable by
TianQin, gLISA, and~by a de-scoped version of gLISA, gLISA$_d$, to~reconstruct the parameters characteristic of a monochromatic
signal.\endnote{gLISA is a constellation of three LISA-like
  satellites in an Earth geosynchronous orbit; gLISA$_d$ differs from
  gLISA by having an acceleration noise level that is worse by three
  orders of magnitude and equal to
  $3.0 \times 10^{-12} \ \rm m/s^2 \ {Hz}^{-1/2}$ over the
  ``mid-region'' frequency band. This acceleration noise level has
  already been demonstrated by the Gravity Recovery and Climate
  Experiment Follow-On (GRACE-FO) mission~\cite{GRACEFO}.} The
mid-band frequency region is expected to contain a wide variety of
sources of sinusoidal signals. The~white-dwarf--white-dwarf binary
systems present in our galaxy and hundreds of thousands to millions of
binary black holes with masses in the (10--100 $\ M_\odot$) range may
be regarded as primary monochromatic sources for these detectors.  The~GW signals emitted by these systems can last for several months in the
mid-band frequency region accessible by these~detectors.

Analyses similar to those presented in this article have already
appeared in the literature for the LISA and TianQin missions
~\cite{Cutler,CutlerVecchio,ChineseTianQinAccuracySinusoids}. There,
however, either the long-wavelength approximation for the detector
response was used over the entire operational frequency band
~\cite{Cutler,CutlerVecchio} or a representation of the interferometer
response in the complex domain~\cite{ChineseTianQinAccuracySinusoids}
resulted in a mathematically incorrect expression of the Doppler
modulation due to the interferometer motion around the~Sun.

Our analysis relies on the published noise spectral densities
characterizing the sensitivities of the TianQin and gLISA missions,
and of the mission concept gLISA$_d$. gLISA, which has been analyzed
for about ten years by a collaboration of scientists and engineers at
the Jet Propulsion Laboratory, Stanford University, the~University of
California San Diego, the~National Institute for Space Research ({INPE,} Brazil), and~Space Systems Loral, was shown to fit the cost limits of
the NASA astrophysics probe class mission program. It is expected to
achieve shot-noise-limited sensitivity in the higher end of its
accessible frequency band as a consequence of its arm length being
equal to roughly $7.4 \times 10^4$~km range, surpassing LISA’s
sensitivity by a factor of about $35$\endnote{The gLISA sensitivity
  and the analysis of the magnitude of the noises that define it have
  been discussed in the main body and appendix of Ref.
~\cite{TAAA}. In~addition, the~LISA Pathfinder experiment~\cite{LPF}
  has demonstrated the noise level of the optical bench adopted by
  LISA to be three orders of magnitude smaller than its anticipated
  value, thereby confirming the scientific capabilities of
  gLISA.}. TianQin and gLISA will reach their optimal sensitivity in a
frequency band that perfectly complements those covered by LISA,
TaiJi, and~advanced LIGO (aLIGO). As~a result, the~combined detection
range for {GWs} will extend across $(10^{-4}$--$10^3)$~Hz
(see Figures below).

Regarding the onboard scientific payload of gLISA, which primarily
includes the laser, optical telescope, and~inertial reference sensor,
we assume a noise performance comparable to that of LISA~\cite{TBDT}. Other subsystems are considered to contribute noise
levels that lead to a high-frequency noise spectrum primarily governed
by photon-counting statistics.  For~further details, we refer the
reader to Appendix A of Ref.~\cite{TAAA}.  As~mentioned earlier, we
will also consider a gLISA de-scoped mission, gLISA$_d$, which differs
from the gLISA specifications by displaying an acceleration noise
level that is worse by three orders of magnitudes. As~gravitational
wave astronomy has now become a reality, it is likely that other
space-based interferometer designs of lower costs and less demanding
technological developments will be pursued.  As~will be shown in this
article, a~mission such as gLISA$_d$ could deliver good science on a
reduced budget as it could rely on a technology that has already been
flown on the Gravity Recovery and Climate Experiment Follow-On
(GRACE-FO) mission~\cite{GRACEFO}\endnote{GRACE-FO demonstrated that two
  Earth-orbiting spacecraft could coherently and continuously track
  each other with laser light and achieve a noise level of
  $1 \ {{\rm nm}/\sqrt{\rm Hz}}$ at frequencies above 100~mHz.}.

The paper is organized as follows. In~Section~\ref{SecII}, we first
derive the expression of the TDI
Michelson response~\cite{TD2020} of a geocentric
interferometer rotating around the Sun and Earth. We carry this out for the
equilateral configurations of the TiaQin, gLISA, and gLISA$_d$
missions. TianQin is a triangular constellation with a nominal arm
length $L = 1.73 \times 10^5$~km, designed to orbit the Earth with a
period $P_s = 2\pi / \omega_s = 3.65$~days while also revolving around
the Sun alongside Earth. The~constellation is inclined at an angle of
$\alpha = 94.7^\circ$ relative to the ecliptic plane.  gLISA and
gLISA$_d$ are instead in a geosynchronous orbit that is $1.5^{\circ}$
inclined with respect to the equator~\cite{TBDT}. These motions
introduce amplitude and Doppler modulations on the observed
monochromatic signals that define the accuracies and precisions of the
measured parameters characterizing them. In~Section~\ref{SecIII}, after~
presenting a brief reminder of the Fisher information matrix formalism, 
we then derive the analytic expressions of the Fisher
information matrix associated with the responses of the three
orthogonal TDI channels $A$, $E$, and~$T$ \cite{PTLA02} to a sinusoidal GW signal\endnote{The
  $A$, $E$, and~$T$ combinations {were first presented in~\cite{PTLA02} and} may be written in terms of the three
  Unequal-Arm Michelson combinations $X$, $Y$, and~$Z$ \cite{TD2020}
  by using the following expressions:
  $A = \frac{(Z - X)}{\sqrt{2}}\ $,
  $E = \frac{(X - 2Y + Z)}{\sqrt{6}} \ $,
  $T = \frac{(X + Y + Z)}{\sqrt{3}} $. {These combinations are designed to maximize the SNR achievable by a space-based interferometer.}}. This is 
described by an amplitude $h_0$, a~frequency $f_{gw}$ in the rest frame of the source, and~two angles ($\theta, \phi$) associated with the location of the source
in the sky.  The~analytic expressions of the Fisher information
matrix, which were derived using the Python library for symbolic
mathematics \texttt{SymPy} \cite{meurer2017sympy} were then imported into
a Python program for graphical representation and analysis. 
In Section~\ref{SecIV}, we discuss the results of our analysis. 
We find that all three missions achieve their best angular source
reconstruction precision in the higher part of their accessible
frequency band, with~an error box better than $10^{-10}$ sr in the
frequency band [$10^{-1}, 10$]~Hz when observing a monochromatic {GW} signal of amplitude $h_0 = 10^{-21}$ and incoming
from a given direction. In~terms of their reconstructed frequencies
and amplitudes, TianQin achieves its best precisions in both
quantities in the frequency band [$10^{-2}, 4 \times 10^{-1}$] Hz,
with a frequency precision $\sigma_{f_{gw}} = 2 \times 10^{-11}$ Hz
and an amplitude precision $\sigma_{h_0} = 2 \times 10^{-24}$. gLISA
matches these precisions in a frequency band slightly higher than that
of TianQin, [$3 \times 10^{-2}, 1$] Hz, as~a consequence of its
shorter arm length. gLISA$_d$, on~the other hand, matches the
performance of gLISA only in the narrower frequency region
[$7 \times 10^{-1}, 1$] Hz, as~a consequence of its higher acceleration
noise at lower~frequencies.

By assuming a signal-to-noise ratio of 10 (averaged over source sky
location and polarization states) for the TianQin and gLISA missions,
we then derive their angular, frequency, and amplitude precisions as
functions of the source sky location for a selected number of {GW} frequencies covering the operational bandwidths of
the three interferometers. Relying on the same GW amplitudes resulting
in an SNR of 10 for gLISA, we then obtain the angular, frequency, and
amplitude precisions for the gLISA$_d$ mission. The~three sets of
results show, for \underline {any} given source-sky location, that
all three missions display a marked precision improvement in the three
reconstructed parameters at higher GW~frequencies.

\section{The Interferometer Response to a Sinusoidal~Signal}
\label{SecII}

\begin{figure}[H]
\includegraphics[width=0.5\textwidth, angle=0.0]{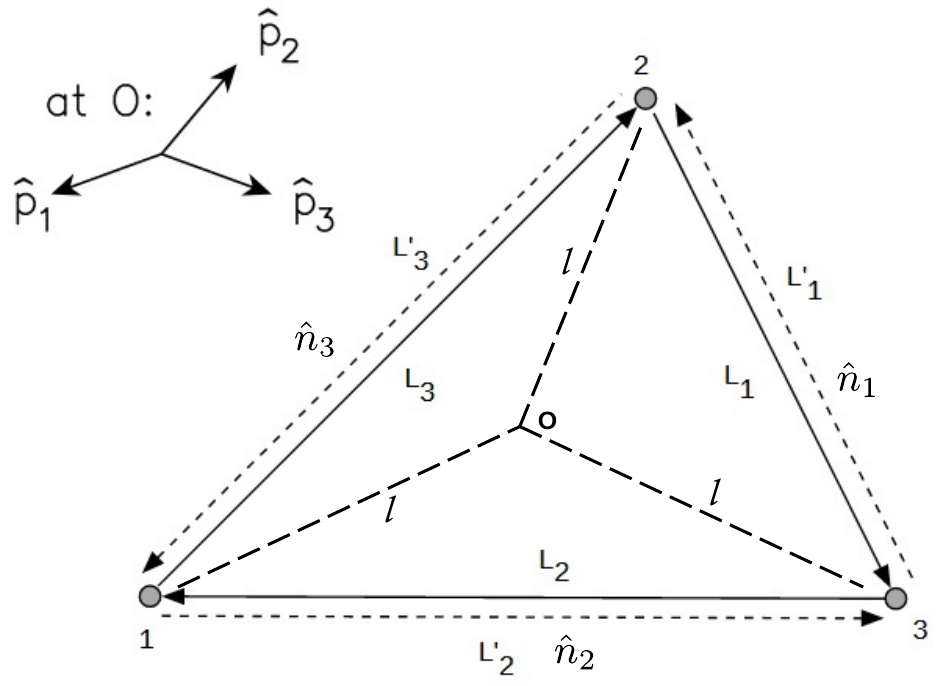}
\caption{Schematic configuration of the interferometer. Each
  spacecraft is positioned at an equal distance from a central
  reference point $o$, with~unit vectors $\hat{p}_{i}$ defining the
  directions from $o$ to each of the three spacecraft. The~unit
  vectors $\hat{n}_{i}$ represent the directional axes between pairs
  of spacecraft, following the specified orientation. The~optical
  paths are denoted by $L_{i}$, with $L{'}_{i}$ depending on whether the
  light beam is seen propagating clock- or anti-clockwise, and the
  spacecraft index $i$ corresponds to the opposite spacecraft. Primed
  and unprimed delays account for differences in the light travel
  times due to the Sagnac effect~\cite{TD2020}.}
    \label{fig:geometry}
\end{figure}

The geometry of the array is shown in Figure~\ref{fig:geometry}. The~three spacecraft continuously exchange six laser beams, with~each
incoming beam being combined with the local laser light at the
receiving optical bench. This process yields six Doppler measurements,
denoted as $y_{ij}$ ($i,j = 1, 2, 3$). To~enable the detection and
analysis of~{GWs} at the expected signal amplitudes, the~frequency fluctuations of the six lasers—present in all Doppler
measurements—must be suppressed to a level below that of secondary
noise sources, such as proof mass and optical path noise~\cite{TD2020}.
\vspace{-6pt}

We adopted the following labeling convention for the Doppler
data. $y_{23}$, for~instance, represents the one-way Doppler shift
recorded at spacecraft $3$ for a signal transmitted from spacecraft
$2$ along arm $1$. $y_{32}$, on~the other hand, denotes the Doppler
shift measured at spacecraft 2 for a signal received from spacecraft 3
along arm $1'$. Due to the rotation of the triangular spacecraft array
around both the Sun and the Earth, the~one-way light travel times
between any pair of spacecraft are generally unequal $L_i \neq L{'}_i$
as a consequence of the Sagnac effect~\cite{S03}. To~accurately
combine the data, it is necessary to account for the signal
propagation delays, which depend on the direction of light travel
along each link. Following~\cite{TEA04}, the~arms are labeled with
single numbers given by the opposite spacecraft; e.g.,~arm $2$ (or
$2'$) is opposite spacecraft $2$; primed delays are used to
distinguish light times taken in the counter-clockwise sense and
unprimed delays for clockwise light times (see
Figure~\ref{fig:geometry}).

Frequency fluctuations arise from various sources, including the
lasers, optical benches, proof masses, fiber optics, and~the inherent
noise at the photodetectors (such as shot-noise fluctuations). These
fluctuations imprint distinct time-dependent signatures on the Doppler
observables; see Refs.~\cite{ETA00,TEA02} for a detailed
discussion. The~one-way Doppler response to {GWs},
denoted as $y^\mathrm{GW}_{ij}$, was initially derived in
Ref.~\cite{AET99} for a stationary spacecraft array and later extended
in Ref.~\cite{KTV04} to account for the realistic orbital
configuration of the LISA array as it orbits the~Sun.

Let us examine, for~example, the~``second-generation'' unequal-arm
Michelson TDI observables~\cite{TDM2022}, denoted as
($X_1, X_2, X_3$). These observables can be expressed in terms of the
Doppler measurements $y_{ij}$ as follows:\endnote{In addition to the
  primary inter-spacecraft Doppler measurements $y_{ij}$
  ($ i, j = 1, 2, 3$), which contain the~{GW} signal,
  each spacecraft also conducts on-board metrology measurements. These
  are necessary due to the presence of two lasers and two proof masses
  in the onboard drag-free control system. However, as~demonstrated in
~\cite{TD2020}, these onboard measurements can be appropriately
  time-delayed and linearly combined with the inter-spacecraft
  measurements, effectively reducing the system to an equivalent
  configuration with only three lasers and six one-way
  inter-spacecraft measurements, simplifying the interferometric
  analysis.}
  \vspace{-2pt}
  {\small
\begin{eqnarray}
X_1 & = & [{(y_{31} + y_{13;2}) + (y_{21} + y_{12;3'})}_{;2'2}
+ {(y_{21} + y_{12;3'})}_{;33'2'2}
+ {(y_{31} + y_{13;2})}_{;33'33'2'2}]
\nonumber
\\
& - &
[(y_{21} + y_{12;3'})
+ {(y_{31} + y_{13;2})}_{;33'}
+ {(y_{31} + y_{13;2})}_{2'233'} +
{(y_{21} + y_{12;3'})}_{;2'22'233'}],
\label{eq:X1}
\end{eqnarray}}with $X_2$ and $X_3$ obtained from Equation~(\ref{eq:X1}) by
appropriately permuting the spacecraft indices.  The~semicolon
notation in Equation~(\ref{eq:X1}) highlights the fact that applying
multiple sequential delays to a measurement is inherently
non-commutative. This arises from the time dependence of the light
travel times $L_i$ and $L{'}_i$ ($i = 1, 2, 3$), meaning that the
order in which delays are applied is crucial for effectively canceling
laser noise~\cite{TEA04,CH03,STEA03,TD2020}. To~be clear, the~delayed
measurement $y_{ij;kl} \equiv y_{ij} (t - L_l (t) - L_k (t - L_l))$ is
generally not equal to $y_{ij;lk}$, illustrating the asymmetry of the
delay operations (using units where the speed of light $c=1$).

It is clear that $X_1$ and the corresponding first generation TDI
combination, $X$, (the unequal-arm Michelson observable valid for a
stationary array~\cite{AET99,ETA00}) will display different responses
to the~{GW} signal and the secondary noise sources.
However, since the corrections introduced by the motion of the array
to the GW signal response and to the secondary noises are proportional
to the product between their time derivatives and the difference
between the actual light travel times and those valid for a stationary
array, it is easy to show~\cite{ETKN2005} that, at~$1$ Hz, the~largest
corrections to the signal and the noises (due to the Sagnac effect)
are about four orders of magnitude smaller than their main
counterparts. Since the amplitudes of these corrections scale linearly
with the Fourier frequency, we can completely disregard this effect
over the entire bands of the interferometers considered~\cite{TEA04}.
Furthermore, for~the orbits of the three arrays analyzed, the~three
arm lengths will differ at most by $\sim$0.2\%~\cite{TAAA} and the
resulting degradation in signal-to-noise ratio introduced by adopting
signal templates that neglect the inequality of the arm lengths will
be of only a few tenths of a percent.  For~these reasons, in~what
follows, we will focus on the expressions of the GW responses of
various second-generation TDI observables by disregarding the
differences in the delay times experienced by light propagating
clockwise and counterclockwise, and~by assuming the three arm lengths
of the considered three geocentric missions to be constant and equal
to their nominal reference values.  In~the case of TianQin, for~example, its arm length
$L= 1.7 \times 10^5 \, \mbox{km} \simeq 0.57 \, \mbox{s}$, while for
gLISA and gLISA$_d$,
$L = 7.4 \ \times 10^4 \, \mbox{km} \simeq 0.25 \, \mbox{s}$. This
approximation has been referred to in the literature as the
rigid adiabatic approximation~\cite{RCP04}, and~the formalism
of Ref.~\cite{Seto04} discussed this for~LISA.

From these considerations, we infer that the expressions of the~{GW} signal and the secondary noises in the
second-generation TDI combinations, ($X_1, X_2, X_3$), can be
expressed in terms of the corresponding equal arm-length combinations,
$(M_1, M_2, M_3$).  For~instance, the~{GW} signal in the
second-generation unequal-arm Michelson combination, $X^{\rm GW}_1$,
can be expressed in terms of the~{GW} response of the
corresponding equal arm-length Michelson combination,
$M^{\rm GW}_1 (t)$, in~the following way~\cite{TL}:
\begin{equation}
X^{\rm GW}_1 (t) = M^{\rm GW}_1 (t) - M^{\rm GW}_1 (t - 2L) - M^{\rm GW}_1 (t - 4L)
+ M^{\rm GW}_1 (t -6L) \ .
\label{X1fromX}
\end{equation}

From Equation~(\ref{X1fromX}) above, we conclude that any data analysis
technique for the second-generation TDI combinations can be obtained
by considering the corresponding equal-arm length TDI expressions. In~what follows, we will focus our attention on the three equal arm-length
Michelson combinations ($M_1, M_2, M_3$).

The expressions of the relative frequency changes
$y_{ij}^{\rm GW}(t)$, induced by a transverse traceless gravitational
wave propagating from the source direction $\hat{k}$, have been
derived in Ref.~\cite{AET99} for a stationary triangular array and are
equal to
\begin{equation}
y_{21}^{\rm GW}(t) = \left [ 1 + \frac{l}{L}(\mu_{1}-\mu_{2}) \right ] [\Psi_{3}(t-\mu_{2}l-L) - \Psi_{3}(t-\mu_{1}l)],
\label{eq:y21}
\end{equation}
\begin{equation}
y_{31}^{\rm GW}(t) = \left [ 1 - \frac{l}{L}(\mu_{3}-\mu_{1}) \right ] [\Psi_{2}(t-\mu_{3}l-L) - \Psi_{2}(t-\mu_{1}l)],
\label{eq:y31}
\end{equation}
where $\mu_{i} l = \hat{k} \cdot \hat{p}_{i} l$ represents the delay
of the gravitational wavefront to the position of the spacecraft
relative to the center of the array and $\hat{p}_{i}$ is the unit
vector indicating the location of spacecraft $i$ from the center $o$
of the array.  The~$\Psi_{i}(t)$ terms contain the effects of the GW
signal at the times of emission and reception of the laser photon
packet and are equal to
\begin{equation}
\Psi_{i}(t) \equiv \frac{\hat{n}_{i} \cdot {\sf H}(t) \cdot \hat{n}_{i}}{2[1-(\hat{k} \cdot \hat{n}_{i})^{2}]},
\label{eq:PSI}
\end{equation}
where ${\sf H}(t) = h_{+}(t) \mathbf{e_{+}} + h_{\times}(t)
\mathbf{e_{\times}}$ is the~{GW} tensor; the three-tensor
$\mathbf{e_{+}}$ and $\mathbf{e_{\times}}$ are traceless and
transverse to $\hat{k}$, and~their components in the TT-gauge
coordinates frame are equal to
\begin{equation}
\mathbf{e_{+}} = 
\begin{pmatrix}
1 & 0 & 0 \\
0 & -1 & 0 \\
0 &  0 & 0 \\
\end{pmatrix}
\hspace{1cm}
\mathbf{e_{\times}} = 
\begin{pmatrix}
0 & 1 & 0 \\
1 & 0 & 0 \\
0 & 0 & 0 \\
\end{pmatrix}.
\label{eq:eplus_cross}
\end{equation}

In Equation~(\ref{eq:PSI}), $h_+(t)$ and $h_\times (t)$ are
the wave's two independent polarization functions. In~the case of a
monochromatic GW signal, they can be expressed as
\begin{eqnarray}
    h_{+}(t) & = & h_{0+} cos(\omega t) \ ,
    \nonumber
    \\
    h_{\times}(t) & = & h_{0 \times} sin(\omega t) \ ,
    \label{h+x}
\end{eqnarray}
where $h_{0+}$ and $h_{0 \times}$ are the GW (constant) amplitudes of
each polarization and $\omega$ is the GW angular frequency.\endnote{The expressions of the wave's two independent amplitudes can
  also be written using complex notation in the following way:
  $h_+ (t) = h_{0+} \ {\exp{[2 \pi i \omega t]} }\ ; \ h_\times = i \
  h_{0\times} \exp{[2 \pi i \omega t]}$. However, care must be taken
  when performing calculations similar to those presented in~\cite{ChineseTianQinAccuracySinusoids} as $\Re$ parts of the
  complex representation of the wave should be taken first.}

The expression of the equal-arm Michelson interferometer $M_{1}(t)$
response to a sinusoidal~{GW} can be written as follows
(see Appendix~\ref{append_A}):
\begin{equation}
M_{1}(t) = h_{0+} \cos{(\omega t)} \, F_{\rm I} + h_{0+} \sin{(\omega t)} \, F_{\rm II},
\label{eq:M1_gnral}
\end{equation}
where 
\begin{align} \label{eq:F_plus}
  F_{\rm I} &= \left ( 1 - \hat{k} \cdot \hat{n}_{3} \right ) [\Psi_{3+} \cos{(\omega \tau_{1})} \nonumber - \mathcal{A} \Psi_{3 \times} \sin{(\omega \tau_{1})}] 
  \nonumber \\
  & \quad - \left ( 1 + \hat{k} \cdot \hat{n}_{2} \right ) [\Psi_{2+} \cos{(\omega \tau_{1})} - \mathcal{A} \Psi_{2 \times} \sin{(\omega \tau_{1})}]  \nonumber \\
            & \quad + \left ( 2\hat{k} \cdot \hat{n}_{3} \right ) [\Psi_{3+} \cos{(\omega \tau_{2})} - \mathcal{A} \Psi_{3 \times} \sin{(\omega \tau_{2})}] \nonumber \\
            & \quad + \left ( 2\hat{k} \cdot \hat{n}_{2} \right ) [\Psi_{2+} \cos{(\omega \tau_{3})} - \mathcal{A} \Psi_{2 \times} \sin{(\omega \tau_{3})}] \nonumber \\
            & \quad - \left ( 1 + \hat{k} \cdot \hat{n}_{3} \right ) [\Psi_{3+} \cos{(\omega \tau_{4})} - \mathcal{A} \Psi_{3 \times} \sin{(\omega \tau_{4})}] \nonumber \\
            & \quad + \left ( 1 - \hat{k} \cdot \hat{n}_{2} \right ) [\Psi_{2+} \cos{(\omega \tau_{4})} - \mathcal{A} \Psi_{2 \times} \sin{(\omega \tau_{4})}], 
\end{align}
and
\begin{align} \label{eq:F_cross}
  F_{\rm II} &= \left ( 1 - \hat{k} \cdot \hat{n}_{3} \right ) [\Psi_{3+} \sin{(\omega \tau_{1})} + \mathcal{A} \Psi_{3 \times} \cos{(\omega \tau_{1})}] 
  \nonumber \\
    & \quad - \left ( 1 + \hat{k} \cdot \hat{n}_{2} \right ) [\Psi_{2+} \sin{(\omega \tau_{1})} + \mathcal{A} \Psi_{2 \times} \cos{(\omega \tau_{1})}]  
\nonumber \\
    & \quad + \left ( 2\hat{k} \cdot \hat{n}_{3} \right ) [\Psi_{3+} \sin{(\omega \tau_{2})} + \mathcal{A} \Psi_{3 \times} \cos{(\omega \tau_{2})}] 
\nonumber \\
    & \quad + \left ( 2\hat{k} \cdot \hat{n}_{2} \right ) [\Psi_{2+} \sin{(\omega \tau_{3})} + \mathcal{A} \Psi_{2 \times} \cos{(\omega \tau_{3})}]  
\nonumber \\
& \quad - \left ( 1 + \hat{k} \cdot \hat{n}_{3} \right ) [\Psi_{3+} \sin{(\omega \tau_{4})} + \mathcal{A} \Psi_{3 \times} \cos{(\omega \tau_{4})}] 
\nonumber \\
& \quad + \left ( 1 - \hat{k} \cdot \hat{n}_{2} \right ) [\Psi_{2+} \sin{(\omega \tau_{4})} + \mathcal{A} \Psi_{2 \times} \cos{(\omega \tau_{4})}]. 
\end{align}

In Equations~(\ref{eq:F_plus}) and (\ref{eq:F_cross}), the delay-times $\tau_{i}$
are equal to $\tau_{1} \equiv (\mu_{1}l+2L)$,
$\tau_{2} \equiv (\mu_{2}l+L)$, $\tau_{3} \equiv (\mu_{3}l+L)$, and
$\tau_{4} \equiv \mu_{1}l = \tau_{1} - 2L$, while
$\mathcal{A} \equiv h_{0 \times} / h_{0+}$.  The~expressions for the
other two equal-arm Michelson interferometers, $M_{2}(t)$ and
$M_{3}(t)$, can be obtained from $M_{1}(t)$ by permutation of the
spacecraft~indices.

Since the array is not stationary, the~{GW} signal will
appear in the equal-arm Michelson measurement as modulated in
amplitude and phase. In~order to derive its expression, it is
convenient to express it in the inertial reference frame centered on
the Solar System Baricenter (SSB). In~the coordinate frame where the
spacecraft are at rest, their positions relative to the center of the
array $\Vec{p}_{i}$, and~the unit vectors along the arms $\hat{n}_{i}$
can be written as follows:
\begin{equation} 
 \Vec{p}_{i} = \frac{L}{\sqrt{3}} ( -\cos2\sigma_i , \sin2\sigma_i, 0),
 \label{eq:ptelesc}
\end{equation} 
and
\begin{equation} 
\hat{n}_{i} = ( \cos\sigma_i , \sin\sigma_i, 0),
\label{eq:ntelesc}
\end{equation} 
where
\begin{equation} 
\label{eq:sigma}
\sigma_i = \frac{3\pi}{2} - \frac{2(i-1)\pi}{3}   .  
\end{equation}

The trajectories of the three GW space observatories analyzed in this
work are geocentric, with~their three spacecraft simultaneously
orbiting Earth and the Sun. Additionally, the~normal vectors of their
detector plane point in specific (constant) directions in the
sky. Since the observatory's guiding centers lie on the ecliptic
plane, it is convenient to introduce a SSB ecliptic coordinate
system. In~these coordinates, we align the $x$ axis with the direction
to the vernal equinox, and~define
$\Vec{r} = R (\cos\eta, \sin\eta, 0)$ as the vector from the origin of
the SSB coordinate system to the guiding center of the array. Here,
$R=1$~AU is constant, the~function $\eta = \Omega t + \eta_0$
describes the motion of the guiding center around the Sun and
$\Omega = 2\pi/1$~yr is the rotation frequency around the Sun. The~vectors $\Vec{p}_{i}$ and $\hat{n}_{i}$ can then be expressed in the
SSB coordinate system as~\cite{KTV04}.\endnote{For ease of
  calculation, we have approximated the trajectory of the centers of
  the arrays around the SSB to be perfectly circular with a period of
  1 year.}
\begin{equation}
  \Vec{\mathfrak{p}}_{i}^{R}(t) = \Vec{r}(t) + \Vec{\mathfrak{p}}_{i}(t) = \Vec{r}(t) + {\sf O}_1 \cdot \Vec{p}_{i},    
\label{eq:pos_SSB}
\end{equation}
and
\begin{equation} 
\hat{\mathfrak{n}}_{i}(t) = {\sf O}_1 \cdot \hat{n}_{i}.
\label{eq:n_SSB}
\end{equation} 

${\sf O}_1$ is the rotation matrix that relates the coordinates
attached to the interferometer to those defined earlier in the SSB,
and is equal to
\begin{equation}
{\sf O}_1 = \left(
\begin{array}{ccc}
\cos\beta & -\sin\beta  & 0 \\
\cos\alpha \, \sin\beta   & \cos\alpha \, \cos\beta & -\sin\alpha \\
\sin\alpha \, \sin\beta   &  \sin\alpha \, \cos\beta &  \cos\alpha \\
\end{array}
\right), 
\label{eq:O1rot}
\end{equation}
where $\alpha$ denotes the inclination of the orbital plane of the
spacecraft array relative to the ecliptic and the function
$\beta = \omega_{s} t + \beta_0$ represents the rotation phase of each
spacecraft around the guiding center of the array with angular
velocity $\omega_{s}$. For~simplicity, we set \mbox{$\eta_0 = \beta_0 = 0$,}
so that at $t=0$, the~vector $\Vec{p}_{1}$ is aligned with the $x$
axis of the SSB coordinate~system.

A transverse traceless tensor, associated with a GW signal emitted by
a source located at latitude $\theta$ and longitude $\phi$ relative to
the SSB coordinate system, is given by the following expression:
\begin{equation}
  \mathbb{H}(t) = {\sf O}^{-1}_2 \cdot {\sf H}(t) \cdot {\sf O}_2 \ ,
\label{eq:H_SSB}
\end{equation}
where 
{\small\begin{equation}
{\sf O}_2 = \left(
\begin{array}{ccc}
\cos\phi\cos\psi -\sin\phi\sin\psi\cos\theta  & \sin\phi\cos\psi +\cos\phi\sin\psi\cos\theta & \sin\psi\sin\theta \\
-\cos\phi\sin\psi -\sin\phi\cos\psi\cos\theta  & -\sin\phi\sin\psi +\cos\phi\cos\psi\cos\theta & \cos\psi\sin\theta \\
\sin\phi\sin\theta                              &  -\cos\phi\sin\theta                         & \cos\theta\\
\end{array}
\right) \ ,
\end{equation}}where ($\theta, \phi, \psi$) are the usual Euler angles for which the
wave's direction of propagation is equal to
$\hat{\texttt{k}} \equiv(\sin\phi\sin\theta, -\cos\phi\sin\theta,
\cos\theta)$.

In the SSB frame and considering that
$\mu_{i} l = \hat{k} \cdot \hat{p}_{i}$, the~terms inside the sine and
cosine functions in Equations~(\ref{eq:F_plus}) and (\ref{eq:F_cross}) assume the following forms:
\begin{align} 
\omega \tau_{1}^{ssb} &= \omega \left ( 2L + \hat{\texttt{k}} \cdot \hat{\mathfrak{p}}_{1} + \hat{\texttt{k}} \cdot \hat{r} \right ) = \omega \mathbb{T}_{1} + \phi_{D}, \label{eq:tau1_SSB} \\[5pt]
\omega \tau_{2}^{ssb} &= \omega \left (L + \hat{\texttt{k}} \cdot \hat{\mathfrak{p}}_{2} + \hat{\texttt{k}} \cdot \hat{r} \right ) = \omega \mathbb{T}_{2} + \phi_{D}, \label{eq:tau2_SSB} \\[5pt]
\omega \tau_{3}^{ssb} &= \omega \left ( L + \hat{\texttt{k}} \cdot \hat{\mathfrak{p}}_{3} + \hat{\texttt{k}} \cdot \hat{r} \right ) = \omega \mathbb{T}_{3} + \phi_{D}, \label{eq:tau3_SSB} \\[5pt]
\omega \tau_{4}^{ssb} &= \omega \left (\hat{\texttt{k}} \cdot \hat{\mathfrak{p}}_{1} + \hat{\texttt{k}} \cdot \hat{r} \right ) = \omega \mathbb{T}_{4} + \phi_{D} \label{eq:tau4_SSB}.
\end{align}

Here, $\mathbb{T}_{a}$ represents the combination of retarded times
$L$ and $(\hat{\texttt{k}} \cdot \hat{\mathfrak{p}}_{i})$, while
\mbox{$\phi_{D} = \omega ( \hat{\texttt{k}} \cdot \hat{r} )$} refers to the
Doppler phase, which can be expressed in terms of the angular
coordinates of the GW source as
$\phi_{D}~=~\omega \, R \, \sin{\theta} \, \sin{(\Omega t - \phi)}$.

Based on these considerations and the coordinate transformations
discussed above, the~interferometer's response to a sinusoidal signal
in the SSB can be written in the following~form:
\begin{equation}
\mathbb{M}_{1} (t) = h_{0+} \cos{(\omega t)} \, \mathbb{F}_{\rm I} + h_{0+} \sin{(\omega t)} \, \mathbb{F}_{\rm II},
\label{eq:M1_ssb}
\end{equation}
with
\begin{align}
\mathbb{F}_{\rm I} &= \left ( H_{\rm I} + \mathcal{A}H_{\rm II} \right ) \cos \phi_{D} - \left ( H_{\rm III} + \mathcal{A}H_{\rm IV} \right ) \sin \phi_{D}, \label{eq:fi_ssb} \\[6pt]
\mathbb{F}_{\rm II} &= \left ( H_{\rm I} + \mathcal{A}H_{\rm II} \right ) \sin \phi_{D} + \left ( H_{\rm III} + \mathcal{A}H_{\rm IV} \right ) \cos \phi_{D}, \label{eq:fii_ssb}
\end{align}
and
\begin{align} \label{eq:Hi_ssb}
H_{\rm I} &= \left ( 1 - \hat{\texttt{k}} \cdot \hat{\mathfrak{n}}_{3} \right ) \{ \Psi_{3+}^{ssb} [\cos{(\omega \mathbb{T}_{1})} -\cos{(\omega \mathbb{T}_{2})} ]\} \nonumber \\
& \quad + \left ( 1 + \hat{\texttt{k}} \cdot \hat{\mathfrak{n}}_{3} \right ) \{ \Psi_{3+}^{ssb} [\cos{(\omega \mathbb{T}_{2})} -\cos{(\omega \mathbb{T}_{4})}]\} \nonumber \\
& \quad - \left ( 1 + \hat{\texttt{k}} \cdot \hat{\mathfrak{n}}_{2} \right ) \{\Psi_{2+}^{ssb} [\cos{(\omega \mathbb{T}_{1})} -\cos{(\omega \mathbb{T}_{3})} ]\} \nonumber \\
& \quad - \left ( 1 - \hat{\texttt{k}} \cdot \hat{\mathfrak{n}}_{2} \right ) \{ \Psi_{2+}^{ssb} [\cos{(\omega \mathbb{T}_{3})} -\cos{(\omega \mathbb{T}_{4})}]\}, 
\end{align}
\begin{align} \label{eq:Hii_ssb}
H_{\rm II} &= -\left ( 1 - \hat{\texttt{k}} \cdot \hat{\mathfrak{n}}_{3} \right ) \{\Psi_{3\times}^{ssb}  [\sin{(\omega \mathbb{T}_{1})} -\sin{(\omega \mathbb{T}_{2})} ]\} \nonumber \\
& \quad - \left ( 1 + \hat{\texttt{k}} \cdot \hat{\mathfrak{n}}_{3} \right ) \{\Psi_{3\times}^{ssb} [\sin{(\omega \mathbb{T}_{2})} -\sin{(\omega \mathbb{T}_{4})}]\} \nonumber \\
& \quad + \left ( 1 + \hat{\texttt{k}} \cdot \hat{\mathfrak{n}}_{2} \right ) \{\Psi_{2\times}^{ssb}  [\sin{(\omega \mathbb{T}_{1})} -\sin{(\omega \mathbb{T}_{3})} ]\} \nonumber \\
& \quad + \left ( 1 - \hat{\texttt{k}} \cdot \hat{\mathfrak{n}}_{2} \right ) \{\Psi_{2\times}^{ssb} [\sin{(\omega \mathbb{T}_{3})} -\sin{(\omega \mathbb{T}_{4})}]\} , 
\end{align}
\begin{align} \label{eq:Hiii_ssb}
H_{\rm III} &= \left ( 1 - \hat{\texttt{k}} \cdot \hat{\mathfrak{n}}_{3} \right ) \{ \Psi_{3+}^{ssb} [\sin{(\omega \mathbb{T}_{1})} -\sin{(\omega \mathbb{T}_{2})} ]\} \nonumber \\
& \quad + \left ( 1 + \hat{\texttt{k}} \cdot \hat{\mathfrak{n}}_{3} \right ) \{ \Psi_{3+}^{ssb} [\sin{(\omega \mathbb{T}_{2})} -\sin{(\omega \mathbb{T}_{4})}]\}  \nonumber \\
& \quad - \left ( 1 + \hat{\texttt{k}} \cdot \hat{\mathfrak{n}}_{2} \right ) \{ \Psi_{2+}^{ssb} [\sin{(\omega \mathbb{T}_{1})} -\sin{(\omega \mathbb{T}_{3})} ]\} \nonumber \\
& \quad - \left ( 1 - \hat{\texttt{k}} \cdot \hat{\mathfrak{n}}_{2} \right ) \{ \Psi_{2+}^{ssb} [\sin{(\omega \mathbb{T}_{3})} -\sin{(\omega \mathbb{T}_{4})}]\}, 
\end{align}
\begin{align} \label{eq:Hiv_ssb}
H_{\rm IV} &= \left ( 1 - \hat{\texttt{k}} \cdot \hat{\mathfrak{n}}_{3} \right ) \{\Psi_{3\times}^{ssb}  [\cos{(\omega \mathbb{T}_{1})} -\cos{(\omega \mathbb{T}_{2})} ]\} \nonumber \\
& \quad + \left ( 1 + \hat{\texttt{k}} \cdot \hat{\mathfrak{n}}_{3} \right ) \{\Psi_{3\times}^{ssb} [\cos{(\omega \mathbb{T}_{2})} -\cos{(\omega \mathbb{T}_{4})}]\} \nonumber \\
& \quad - \left ( 1 + \hat{\texttt{k}} \cdot \hat{\mathfrak{n}}_{2} \right ) \{\Psi_{2\times}^{ssb}  [\cos{(\omega \mathbb{T}_{1})} -\cos{(\omega \mathbb{T}_{3})} ]\} \nonumber \\
& \quad - \left ( 1 - \hat{\texttt{k}} \cdot \hat{\mathfrak{n}}_{2} \right ) \{\Psi_{2\times}^{ssb} [\cos{(\omega \mathbb{T}_{3})} -\cos{(\omega \mathbb{T}_{4})}]\}.
\end{align}

After substituting Equations~(\ref{eq:fi_ssb}) and (\ref{eq:fii_ssb}) into Equation~(\ref{eq:M1_ssb}) and some straightforward algebra, $\mathbb{M}_{1}(t)$ assumes the following form
\begin{equation}
\mathbb{M}_{1}(t) = h_{0+} \left ( H_{\rm I} + \mathcal{A}H_{\rm II} \right ) \cos{\Phi} + h_{0+} \left ( H_{\rm III} + \mathcal{A}H_{\rm IV} \right ) \sin{\Phi}, 
\label{eq:M1_ssb1}
\end{equation}
where
$\Phi \equiv \omega t - \phi_{D} = \omega t - \omega \, R \,
\sin{\theta} \, \sin{(\Omega t - \phi)}$ is the GW phase relative to the SSB.
Applying a procedure analogous to that described above, we can
derive the responses of the other two Michelson interferometers
$\mathbb{M}_{2} (t)$ and $\mathbb{M}_{3} (t)$  relative to the~SSB.

Furthermore, we can obtain the responses of the three orthogonal
TDI channels $A$, $E$, and~$T$ in terms of $\mathbb{M}_{1}$,
$\mathbb{M}_{2}$, and~$\mathbb{M}_{3}$ by relying on the following expressions:
\begin{eqnarray}
A &=& \frac{(\mathbb{M}_{3} - \mathbb{M}_{1})}{\sqrt{2}}, \\
E &=& \frac{(\mathbb{M}_{1} - 2\mathbb{M}_{2} + \mathbb{M}_{3})}{\sqrt{6}}, \\
T &=& \frac{(\mathbb{M}_{1} + \mathbb{M}_{2} + \mathbb{M}_{3})}{\sqrt{3}}.
\end{eqnarray}

After some algebra, the three orthogonal channels ($A, E, T$) assume the following forms:
\begin{align}
A &= h_{0+} \left [ \left ( H_{\rm I(A)} + \mathcal{A}H_{\rm II(A)} \right )\cos{\Phi} + \left ( H_{\rm III (A)} + \mathcal{A}H_{\rm IV (A)} \right ) \sin{\Phi}  \right ], \label{eq:TDI_ssb_A} \\[8pt]
E &= h_{0+} \left [ \left ( H_{\rm I(E)} + \mathcal{A}H_{\rm II(E)} \right )\cos{\Phi} + \left ( H_{\rm III (E)} + \mathcal{A}H_{\rm IV (E)} \right ) \sin{\Phi}  \right ],  \label{eq:TDI_ssb_E} \\[8pt] 
T &= h_{0+} \left [ \left ( H_{\rm I(T)} + \mathcal{A}H_{\rm II(T)} \right )\cos{\Phi} + \left ( H_{\rm III (T)} + \mathcal{A}H_{\rm IV (T)} \right ) \sin{\Phi}  \right ],  \label{eq:TDI_ssb_T}
%
\end{align}
where
\begin{eqnarray}
H_{\chi (A)} & \equiv & \frac{H_{\chi (3)} - H_{\chi (1)}}{\sqrt{2}} \ ,
\nonumber \\
H_{\chi (E)} & \equiv & \frac{H_{\chi (1)} - 2H_{\chi (2)} + H_{\chi (3)}}{\sqrt{6}} \ ,
\nonumber \\
H_{\chi (T)} & \equiv & \frac{H_{\chi (1)} + H_{\chi (2)} + H_{\chi (3)}}{\sqrt{3}} \ ,
\label{eq:TDI_H}
\end{eqnarray}
with the index $\chi = {\rm I,II,III,IV}$. The~terms
$H_{\chi (i)} \ , \ i=1, 2, 3$ are the functions derived earlier
defining the expressions of the responses
$\mathbb{M}_{i} \ , \ i=1, 2, 3$.

\section{The Fisher Information Matrix~Formalism}
\label{SecIII}

Different sources of~{GWs} emit distinct types of
signal, which are characterized by properties intrinsically linked to
their physical parameters.  These may be the distribution of the
source mass, its distance to the interferometer, its location in the
sky, and~the angular frequency of the emitted radiation. To~understand
the physical nature of the source that emitted an observed GW signal,
it is essential to estimate the parameters that characterize it and
evaluate their precisions. Here, we will estimate the precisions
achieved by the~{GW} missions TianQin, gLISA, and
gLISA$_d$ by relying on the Fisher Information Matrix (FIM) formalism
in the case of sinusoidal signals. We will assume these signals to be
characterized by an amplitude $h_0$, a~frequency
$f_{gw} = \omega/{2 \pi}$, and~two Euler angles ($\theta, \phi$)
associated with the wave's direction of propagation. As~we will
describe in more detail in the section presenting our results, we have
limited our analysis to linearly and circularly polarized signals.
This is because the results corresponding to an arbitrary polarization
will be ``in between'' those~presented.

The Fisher information matrix of a~{GW} interferometer
response $M(t)$, whose Fourier transform is denoted by $\tilde{M}(f)$,
is given by the following general expression (see, \mbox{e.g., {\cite{Cutler,Seto_2002,Maggiore:2007ulw}}):}

\begin{equation}
  \Gamma^M_{ij} = 4 \Re \int_{0}^{\infty} \frac{\partial_{i} \tilde{M}(f) \partial_{j} \tilde{M}^*(f)}{S_{n}^M(f)} df \ ,
\label{eq:fisher_matrix_fr_domain}
\end{equation}
where $S_{n}^M$ represents the one-sided noise power spectral density
and $\partial_i M \equiv \partial M / \partial\lambda_i$ is the
partial derivative of the interferometer response to a gravitational
wave signal with respect to the component $\lambda^{i}$ of the
parameter vector
$\boldsymbol{\lambda} \equiv (h_0, f_{gw}, \theta , \phi)$. Since the
noise spectrum can be treated as constant over the relatively narrow
bandwidth centered on the frequency $f_{gw}$ of the signal,
Equation~(\ref{eq:fisher_matrix_fr_domain}) can be rewritten in the
following form as a consequence of the Parseval theorem (see, e.g., {\cite{Takahashi_Seto2002, ChineseTianQinAccuracySinusoids}}):
\begin{equation}
\Gamma^M_{ij} = \frac{4}{S_{n}^M(f_{gw})} \int_{0}^{\infty}
\partial_{i} M(t) \partial_{j} M(t) dt .
\label{eq:fisher_matrix_time_domain}
\end{equation}

Equation~(\ref{eq:fisher_matrix_time_domain}) allows us to derive the
following expressions for the Fisher information matrices of the
optimal combinations ($A, E, T$)~{\cite{PTLA02, TD2020, KTV04}}:

\begin{subequations}
\begin{align}
\Gamma^A_{ij} &= \frac{4}{S_{n}^A(f_{gw})} \int_{0}^{\infty} \partial_{i} A(t) \partial_{j} A(t) dt, \\[8pt]
\Gamma^E_{ij} &= \frac{4}{S_{n}^E(f_{gw})} \int_{0}^{\infty} \partial_{i} E(t) \partial_{j} E(t) dt, \\[8pt]
\Gamma^T_{ij} &= \frac{4}{S_{n}^T(f_{gw})} \int_{0}^{\infty}
                \partial_{i} T(t) \partial_{j} T(t) dt ,
\end{align}
\end{subequations}
where $S_{n}^A$, $S_{n}^E$, and $S_{n}^T$ are the one-sided noise power
spectral densities of the ($A, E, T$) combinations, respectively.

Under the assumption of Gaussian noise, from~the above expressions, it
is easy to see that the Fisher information matrix for the combined
($A, E, T$) configuration, $\Gamma_{ij}$, is equal to the sum of
Fisher information matrices of the optimal combinations ($A, E, T$):
\begin{equation} 
\Gamma_{ij} = \Gamma^A_{ij} + \Gamma^E_{ij} + \Gamma^T_{ij}. 
\label{eq:final_fisher_matrix}
\end{equation}

Our analysis of the parameter precisions achievable by the three space
missions will therefore rely on the expression of the Fisher
information matrix obtained from
\mbox{Equation~(\ref{eq:final_fisher_matrix}).}

As mentioned earlier, we limited our analysis to GW signals
characterized by (i)~linear polarization, for~which
$h_{0+} \equiv h_0$ and $h_{0\times} = 0$, and~(ii) circular
polarization, with \mbox{$h_{0+} = h_{0\times} \equiv h_0$.} In~the case of
binary systems, for~instance, these scenarios correspond to specific
orientations of the source orbital plane relative to the line of sight
to the detector: edge-on, where the orbital plane is aligned with the
line of sight, and~face-on, where the orbital plane is perpendicular
to the line of sight. For~this reason, our Fisher information matrix
has dimensions $4\times4$ in the parameter~space.

Since the inverse of the Fisher information matrix is equal to the
covariance matrix, we conclude that the parameters' precisions are
equal to~ the following \cite{Maggiore:2007ulw}:
\begin{equation}
  \left \langle \Delta \lambda_{i} \Delta \lambda_{j}\right \rangle = (\Gamma^{-1})_{ij} \equiv \sigma_{ij}.
\end{equation}

Note that the diagonal elements of the above matrix represent the
variances of the corresponding parameters, while the off-diagonal ones
are the covariances between pairs of them. To~quantify the error box
$\Delta \Omega$ of source sky localization, we will use the following
estimate of the ellipse area determined by the errors
$\sigma_{\theta \theta}$, $\sigma_{\phi \phi}$, and~$\sigma_{\theta \phi}$,
\begin{equation}
\label{eq:error-box}
\Delta \Omega \equiv \pi \sqrt{\sigma_{\theta\theta}\sigma_{\phi\phi} - \sigma_{\theta\phi}^2}.
\end{equation}

\subsection*{Expressions of the Signal's~Derivatives}

To derive the expression of the Fisher information matrix (Equation
(\ref{eq:fisher_matrix_time_domain})), we need to calculate the
derivatives of the detector's TDI responses to the signal with respect
to the parameters $\lambda_i \ \ i = 1, 2, 3, 4$.  They can be
obtained from Equations~(\ref{eq:TDI_ssb_A})--(\ref{eq:TDI_ssb_T}). The~derivation of the response of the $A$
combination is shown below. The~expressions for the other two TDI
responses, $E$ and $T$, follow a similar procedure and
structure. Thus, after~some long but straightforward algebra, the~derivatives of $A$ can be expressed in the following form:
\begin{equation}
    \partial_i  A = D_{1i} \, \cos \Phi + D_{2i} \, \sin \Phi, 
\end{equation}
where
\begin{align} 
D_{1i} &= \frac{1}{\sqrt{2}} \left [ \partial_i (h_{0+} H_{\rm I (A)}) + \partial_i (h_{0+} \mathcal{A} H_{\rm II(A)}) + h_{0+} H_{\rm III(A)} \partial_i\Phi + h_{0+} \mathcal{A} H_{\rm IV (A)} \partial_i\Phi \right ], \\[6pt] 
D_{2i} &= \frac{1}{\sqrt{2}} \left [ \partial_i (h_{0+} H_{\rm III(A)}) + \partial_i (h_{0+} \mathcal{A} H_{\rm IV(A)}) - h_{0+} H_{\rm I(A)} \partial_i\Phi - h_{0+} \mathcal{A} H_{\rm II(A)} \partial_i\Phi \right ]  \ ,
\label{eq:D12_ssb}
\end{align}
and the expressions of $H_{\chi (A)}$ are given by
Equation~(\ref{eq:TDI_H}).
From these expressions, we then obtain the product of the two derivative terms as follows:
\begin{equation}
\partial_i  A \, \partial_j  A  =  D_{1i} D_{1j} \cos^{2}\Phi + D_{2i} D_{2j} \sin^{2}\Phi + (D_{1i} D_{2j} + D_{2i} D_{1j}) \cos\Phi \sin\Phi   \label{eq:DMij_ssb} \ .  \vspace{1pt}  
\end{equation}

Note that ($D_{1i} , D_{2i}$) are functions of the parameter vector $\boldsymbol {\lambda}$, and~both depend on the
derivatives with respect to $\lambda_i$.  Furthermore, by~applying the
trigonometric identities $\cos^{2}\Phi = [1 + \cos(2\Phi)]/2$, $\sin^{2}\Phi = [1 - \cos(2\Phi)]/2$, and $\cos\Phi \sin\Phi = [\sin{(2\Phi)}]/2$, Equation~(\ref{eq:DMij_ssb}) can be rewritten as follows:
\begin{align}
\partial_i  A \, \partial_j  A &= \frac{1}{2} \left[ D_{1i}D_{1j} + D_{2i}D_{2j} \right] + 
\frac{1}{2} \left[ D_{1i}D_{1j} - D_{2i}D_{2j} \right] \cos(2\Phi) \nonumber \\
&\quad + \frac{1}{2} \left[ D_{1i} D_{2j} + D_{2i} D_{1j} \right] \sin(2\Phi).
\label{eq:final_mult_der_terms}
\end{align}

The integral appearing in the Fisher matrix , with~the integrand given by Equation~(\ref{eq:final_mult_der_terms}), is evaluated over an assumed observation time $T_{obs}$ equal to one year. During~such a period of time, the~terms in the integrand multiplying 
$\cos 2\Phi$, and~$\sin 2\Phi$ vanish. This is
because their time dependencies are periodic with periods much shorter
than one year and are therefore orthogonal to the functions
$\cos 2\Phi$ and~$\sin 2\Phi$. For~this reason, the~expression of the element $i,j$ in the Fisher information matrix integrand reduces to the following one:
\begin{equation}
    \partial_i  A \, \partial_j  A = \frac{1}{2}\left(D_{1i}D_{1j} + D_{2i}D_{2j}\right) .
    \label{eq:simp_mul_der_terms}
\end{equation}

The analytic expressions of the partial derivatives of the
interferometer response were derived using the symbolic package {\it
  SymPy} \cite{meurer2017sympy}. A~specialized function was developed
to systematically identify all time-dependent terms within the
equations. These terms were organized into a dataframe with one column
for the time-dependent components, another column for their
corresponding coefficients, and~a third for the expressions of their
symbolic integrals. The~final expression of the integral was obtained
by summing all the coefficients in the dataframe, each multiplied by
its corresponding symbolic integral. This approach was essential to
avoid potential errors associated with numerical integration and to
obtain the final elements of the Fisher information matrix as
functions of the GW~parameters.

\section{Parameters~Precisions}
\label{SecIV}

In this section, we evaluate the measurement precisions of the
parameters ($h_0, f_{gw}, \theta, \phi $) that characterize a
monochromatic~{GW} signal. We derive their magnitudes
for (i) linearly and (ii) circularly polarized signals, in~terms of
the location of the source in the sky described by angles ($\theta$,
$\phi$), the~signal frequency $f_{gw}$, and~for a selected number of
GW amplitudes $h_0$. The~precisions of the angular parameters
($\theta, \phi$) are combined in an angular error box $\Delta \Omega$
whose size is determined through Equation~(\ref{eq:error-box}).

\begin{figure}[H]
\includegraphics[width=0.8\textwidth, angle=0.0]{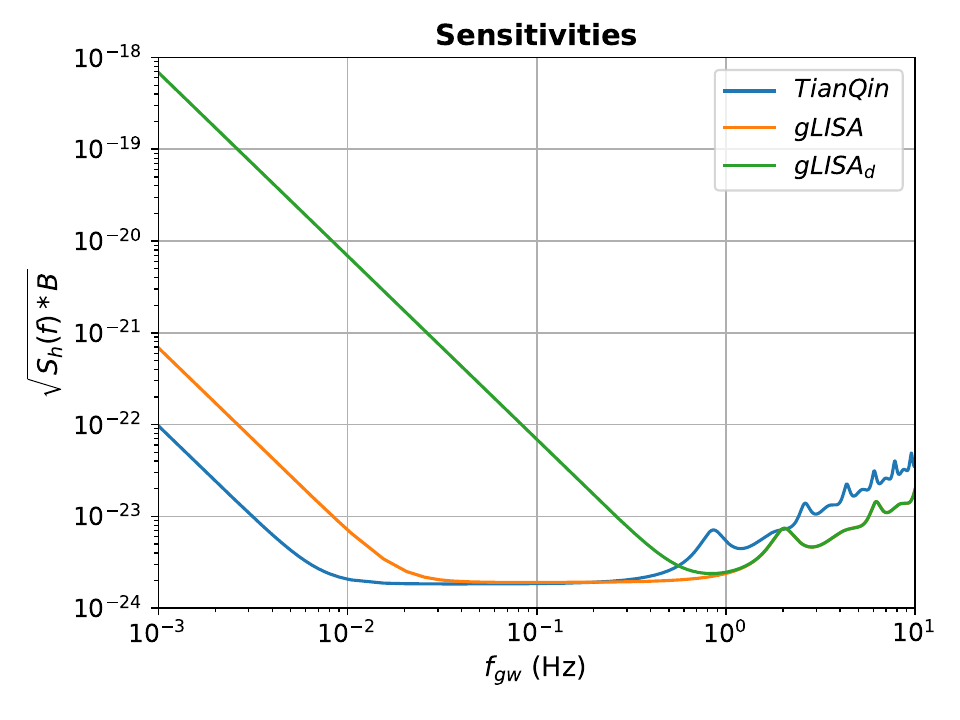}\vspace{-3pt}
\caption{Sensitivity curves, $\sqrt{S_h(f) \ * \ B}$, for~the three
  missions considered~\cite{TD2020}. $S_h(f)$ is equal to the noise
  spectrum divided by the r.m.s. GW response taken over sources
  randomly distributed over the celestial sphere and of random
  polarization states. An~integration time of one year has been
  assumed, which corresponds to a frequency bin
  $B = 3.3 \ \times 10^{-8} \ {\rm Hz}$. }
    \label{fig:sensitivities}
\end{figure}

Our results are obtained by integrating the Fisher information matrix
over a period $T_{obs}=1$~year. The~GW amplitudes are selected by
requiring the average SNR of the TianQin and gLISA missions to be
equal to 10 at the following selected frequencies:
\mbox{$f_{gw} = 10^{-3}, 10^{-2}, 10^{-1}, 1, 10$~Hz.} From~the graph of the
optimal sensitivity~\cite{TD2020} of the gLISA mission (see
Figure~\ref{fig:sensitivities}), we can then infer the values of the GW
amplitudes that correspond to an SNR of 10 and use them in the analysis
of gLISA$_d$. As~we will see, since gLISA$_d$ is penalized at lower
frequencies by an acceleration noise that is 1000 times larger than
that of gLISA, its scientific capabilities are severely impacted in
this frequency band. It should be noted, however, that since gLISA and
gLISA$_d$ share the same trajectory, their precisions would become
equal with GW signals of larger amplitudes, resulting in an SNR of 10 in
gLISA$_d$.

Our analysis is performed using optimal combinations ($A, E, T$)
for an equal-arm array. Their corresponding one-sided noise power
spectral densities ($S_n^A (f), S_n^E(f), S_n^T(f)$) are given by the
following expressions~\cite{TD2020}:

\begin{align}
  S_n^A (f) = S_n^E(f) &= 8 \cdot S_{\text{acc}} \left[ 1 + \cos(2\pi f L) + \cos^2(2\pi f L) \right] 
                  + 2 \cdot S_{\text{opt}} \left[ 2 + \cos(2\pi f L) \right], \\
  S_n^T(f)  &= 8 \cdot \sin^2(\pi f L) \left[ 4 \cdot \sin^2(\pi f L) \cdot S_{\text{acc}} + S_{\text{opt}} \right] \ ,
\end{align}
where $S_{\text{acc}} (f)$ and $S_{\text{opt}} (f)$ are the
acceleration and optical path noise spectra, respectively~\cite{TD2020}. The~magnitude of these spectra depends on the specific
GW mission considered, and~they are equal to
\begin{eqnarray}
    S^{\text{op}}_{\text{TianQin}}(f) &=& 4.0 \times 10^{-40} \cdot f{}^2 \, \text{Hz}^{-1/2}, \\
    S^{\text{pm}}_{\text{TianQin}}(f) &=& 2.8 \times 10^{-49} \cdot f{}^{-2} \, \text{Hz}^{-1/2}, \\
    S^{\text{op}}_{\text{gLISA}}(f) &=& 7.7 \times 10^{-41} \cdot f^2 \, \text{Hz}^{-1/2}, \\
    S^{\text{pm}}_{\text{gLISA}} (f) &=& 2.5 \times 10^{-48} \cdot f^{-2} \, \text{Hz}^{-1/2} \\
    S^{\text{op}}_{\text{gLISA}_d} (f) &=& 7.7 \times 10^{-41} \cdot f^2 \, \text{Hz}^{-1/2}, \\
    S^{\text{pm}}_{\text{gLISA}_d} (f) &=& 2.5 \times 10^{-42} \cdot f^{-2} \, \text{Hz}^{-1/2} \ .
\end{eqnarray}

The above expressions of the TianQin noise spectra were obtained from~\cite{ChineseTianQinAccuracySinusoids}, those for gLISA are as in~\cite{TBDT}, and~those for gLISA$_d$ differ from those of gLISA
by degrading the magnitude of its acceleration noise by a factor of
$10^3$. Note that these noise spectra are for relative frequency
fluctuations as we work with fractional Doppler measurements
throughout this~article.

Based on these noise spectra and the expression of the Fisher
information matrix derived earlier, we can then derive the parameter
precisions characterizing a monochromatic GW signal. In~the following
subsections, we present our results for the three interferometer
missions. We first plot the source location error, $\Delta \Omega$, as~a function of the GW frequency and for three values of the GW
amplitude: $h_0=10^{-23} \ , \ h_0=10^{-22} \ , \ h_0=10^{-21}$. This
is achieved for both linearly and circularly polarized GW signals to
quantify the differences between them. We then plot the precisions of
the GW amplitude $\sigma_{h_0}$ and the GW frequency $\sigma_{f_{gw}}$
as functions of the GW frequency $f_{gw}$ and polarization states. Our
results agree quite well with the corresponding analytic expressions
($\delta \Omega, \delta f_{gw}, \delta h_0/h_0$) given below~\cite{Seto_2002,Takahashi_Seto2002}:
\begin{eqnarray}
  \delta \Omega & = & \frac{2}{\pi R^2} \ {(SNR \times f_{gw})}^{-2} 
  \label{DeltaOmega} \ , 
  \\
  \delta_{f_{gw}} & = & \frac{4 \sqrt{3}}{\pi} \ {(SNR \times T_{obs}})^{-1}
  \label{Deltaf} \ ,
  \\
  \frac{\delta_{h_0}}{h_0} & = & ({SNR})^{-1} \ ,
  \label{Deltah}
\end{eqnarray}
where $R = 1 \ {\rm AU}$ is the nominal distance from the center of
the interferometer to the SSB (in seconds), $T_{obs}$ is the
observation time, and $SNR$ is the corresponding signal-to-noise ratio
averaged over the source directions and polarization states of the
wave.
 
The presented analysis offers insight into the relative performance of
the detectors. We consider signals at frequencies
$f_{gw} = 10^{-3}, 10^{-2}, 10^{-1}, 1, 10$ Hz and whose amplitudes
are such as to result in an average SNR of 10 for the TianQin and
gLISA missions. In~Table~\ref{tab:snr_values}, we provide the GW
amplitudes that result in such an SNR for TianQin and gLISA, and~we
provide the average SNR of gLISA$_d$ when the values for the
amplitudes are as in the case of gLISA. Since the SNR achievable by
gLISA$_d$ with these amplitudes at frequencies smaller than $1$~Hz is
less than $1$, it will come as no surprise that its achievable
precisions in the signal parameters will be rather poor in this part
of the band. However, at~frequencies larger than $1$~Hz, gLISA$_d$ will
equal the performance of~gLISA.

\begin{table}[H]
\caption{Values of the GW amplitudes, at~the selected five GW
  frequencies, resulting in an average SNR of 10 for TianQin and~  gLISA. The~SNRs for gLISA$_d$ at the gLISA GW amplitudes are also~provided.}
\setlength{\tabcolsep}{8.3mm}
\resizebox{\linewidth}{!}{\begin{tabular}{cccc}
\toprule
\textbf{Frequency (Hz)} & \textbf{\textbf{TianQin (}\boldmath{$h_0$}\textbf{)}} & \textbf{\textbf{gLISA (}\boldmath{$h_0$}\textbf{)}} & \textbf{\textbf{gLISA}\boldmath{$_d$} \textbf{(SNR)}} \\
\midrule 
$10^{-3}$    & $9.68 \times 10^{-22}$ & $6.86 \times 10^{-21}$ & 0.01  \\
$10^{-2}$    & $2.07 \times 10^{-23}$ & $7.11 \times 10^{-23}$ & 0.01  \\
$10^{-1}$    & $1.85 \times 10^{-23}$ & $1.90 \times 10^{-23}$ & 0.28  \\
$1$          & $5.37 \times 10^{-23}$ & $2.38 \times 10^{-23}$ & 9.68  \\
$10$         & $3.48 \times 10^{-22}$ & $1.99 \times 10^{-22}$ & 10.00  \\
\bottomrule
\end{tabular}}
\label{tab:snr_values}
\end{table}
\unskip

\subsection{\texorpdfstring{TianQin Parameter Estimation~Errors}{TianQin Parameter Estimation Errors}}

\begin{figure}[H]
\includegraphics[width=0.8\textwidth, angle=0.0]{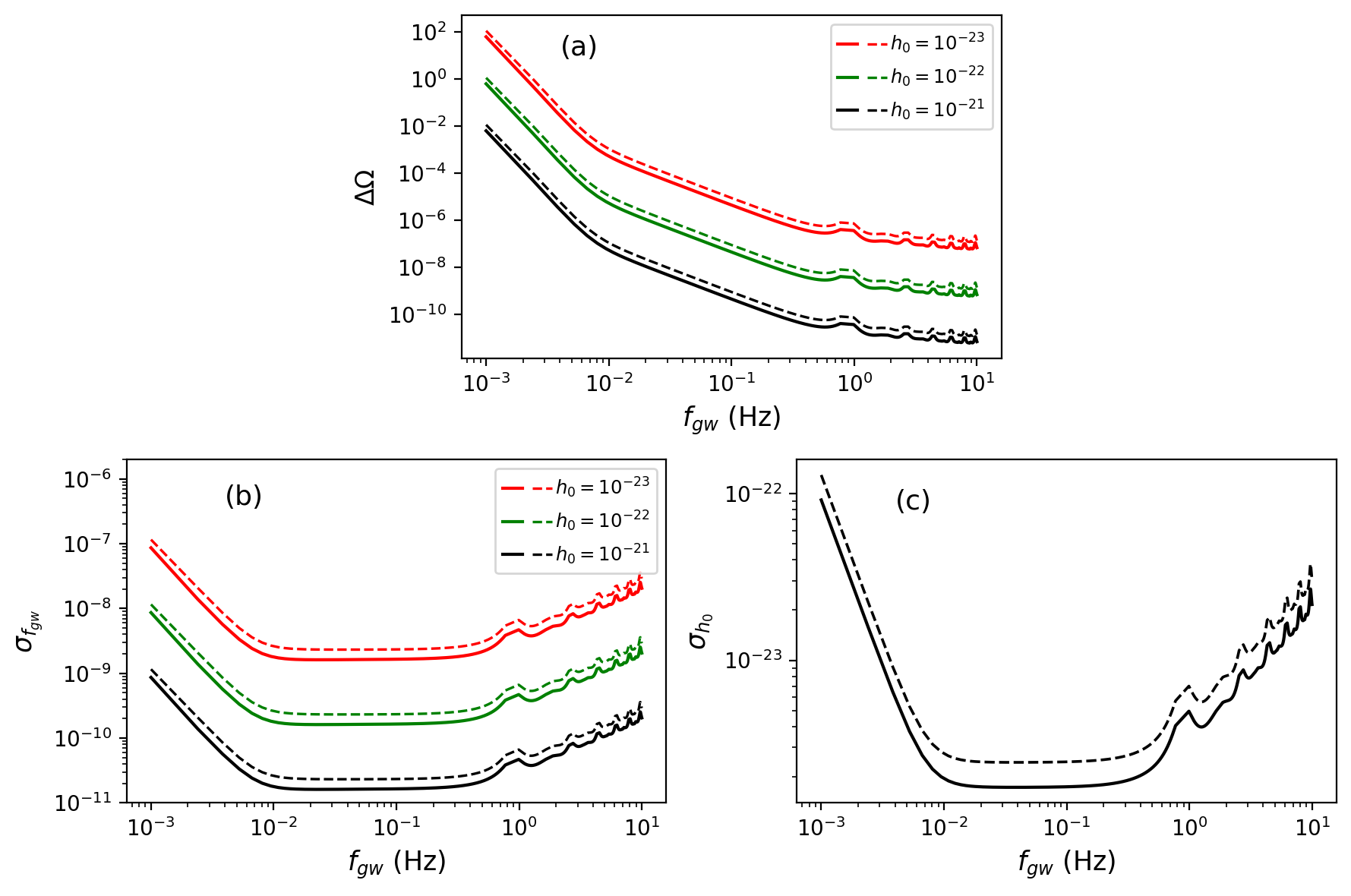}
\caption{TianQin angular $\Delta \Omega$ (\textbf{a}), frequency 
  $\sigma_{f_{gw}}$ (\textbf{b}), and amplitude $\sigma_{h_0}$ (\textbf{c}) precisions as
  functions of the~{GW} frequency, $f_{gw}$, and~for
  three values of the~{GW} amplitude. The~source location was selected to be at ($\theta = \pi/5$, $\phi = 4 \pi/3$), which corresponds to the sky location of a galactic binary system to be observed by TianQin. Continuous lines are for circular polarization, while dashed lines are for linear polarization. See text for a detailed discussion.}
\label{TQ_AngRes_VS_fre}
\end{figure}

In Figure~\ref{TQ_AngRes_VS_fre}a, we plot the angular resolution,
$\Delta \Omega$, of~TianQin as a function of the signal frequency
$f_{gw}$ and for three values of the GW amplitude
($h_0=10^{-23} \ , \ h_0=10^{-22}$, \mbox{$h_0=10^{-21}$)}.  This is achieved
by selecting the source location to be at ($\theta = \pi/5$,
$\phi = 4 \pi/3$), which corresponds to the sky location of a galactic
binary system to be observed by TianQin~\cite{TianQin}. The~effects of
GW polarization are also investigated by plotting the angular
resolutions for linear (dashed lines) and circular (continuous lines)
polarizations. As~expected, the~angular errors for the
two polarizations differ by about a factor of $2$, while their values
scale quadratically with the wave's amplitude
~\cite{Bretthorst1988Bayesian} (see Equation~(\ref{DeltaOmega})). Also, the~angular errors given by Equation~(\ref{DeltaOmega}) and our results
obtained using the Fisher information matrix formalism are in good
agreement, as~can easily be verified. In~Figure~\ref{TQ_AngRes_VS_fre}b, we then plot the precision of the~{GW} frequency, $\sigma_{f_{gw}}$, as~a function of the
GW frequency and the same three GW amplitudes. We may notice, in~agreement with Equation~(\ref{Deltaf}), that it scales linearly with the GW
amplitude, and~the results for the two polarizations differ only by a
factor of $\sqrt{2}$. In~Figure~\ref{TQ_AngRes_VS_fre}c, we then
show the precision of the~{GW} amplitude,
$\sigma_{h_0}$. Again, in~agreement with Equation~(\ref{Deltah}), we see
that it is independent of the value of the GW amplitude itself and
depends mildly on the polarization state of the wave, as~the dashed
and continuous lines again differ by a factor of $\sqrt{2}$.

{It is important to highlight that the frequency and amplitude errors are proportional to the sensitivity (Figure~\ref{fig:sensitivities}), exhibiting larger values at both low and high frequencies. In~contrast, the~angular error decreases as frequency increases, achieving its best value around $10$~Hz---a characteristic that does not align with the sensitivity curve. This behavior can be understood through the frequency dependence of the angular precision. Equations~(\ref{DeltaOmega})--(\ref{Deltah}) describe the general relationship between the precision of the observables, the~SNR, and~the GW frequency. In~particular, for~a given GW amplitude, the~SNR is inversely proportional to the sensitivity curve. Consequently, Equations~(\ref{Deltaf}) and~(\ref{Deltah}), which determine the precision of the GW frequency and amplitude, reflect their proportionalities to the sensitivity curve. Equation~(\ref{DeltaOmega}), on~the other hand, shows that 
the angular precision is proportional to the squared sensitivity and inversely proportional to the square of the GW frequency. This results in lower precision at lower frequencies compared to higher ones, as seen in part (a) of Figure~\ref{TQ_AngRes_VS_fre}.} 

\begin{figure}[H]
\subfloat[\centering]{\includegraphics[width=0.8\textwidth]{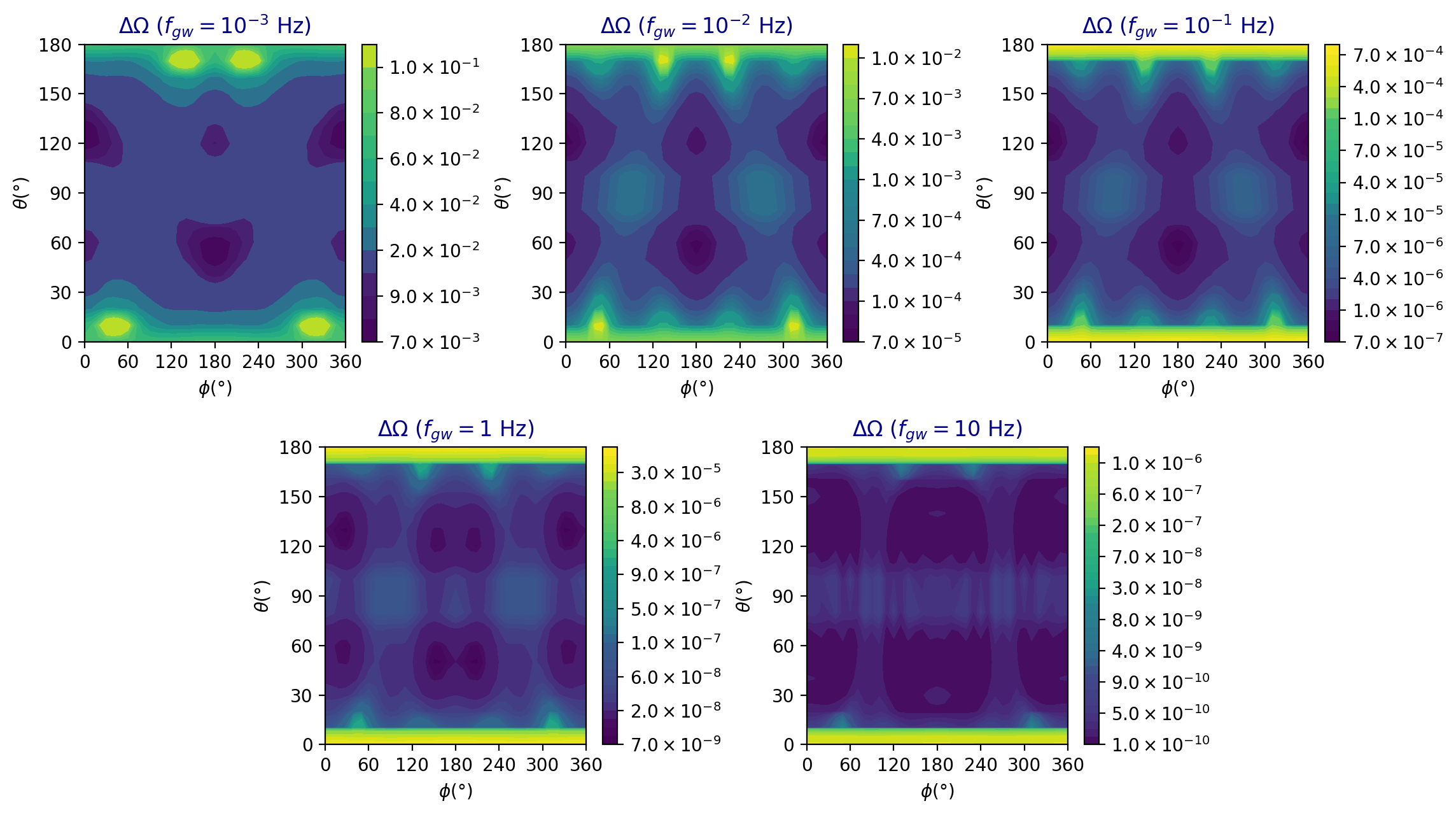}} \\

\subfloat[\centering]{\includegraphics[width=0.8\textwidth]{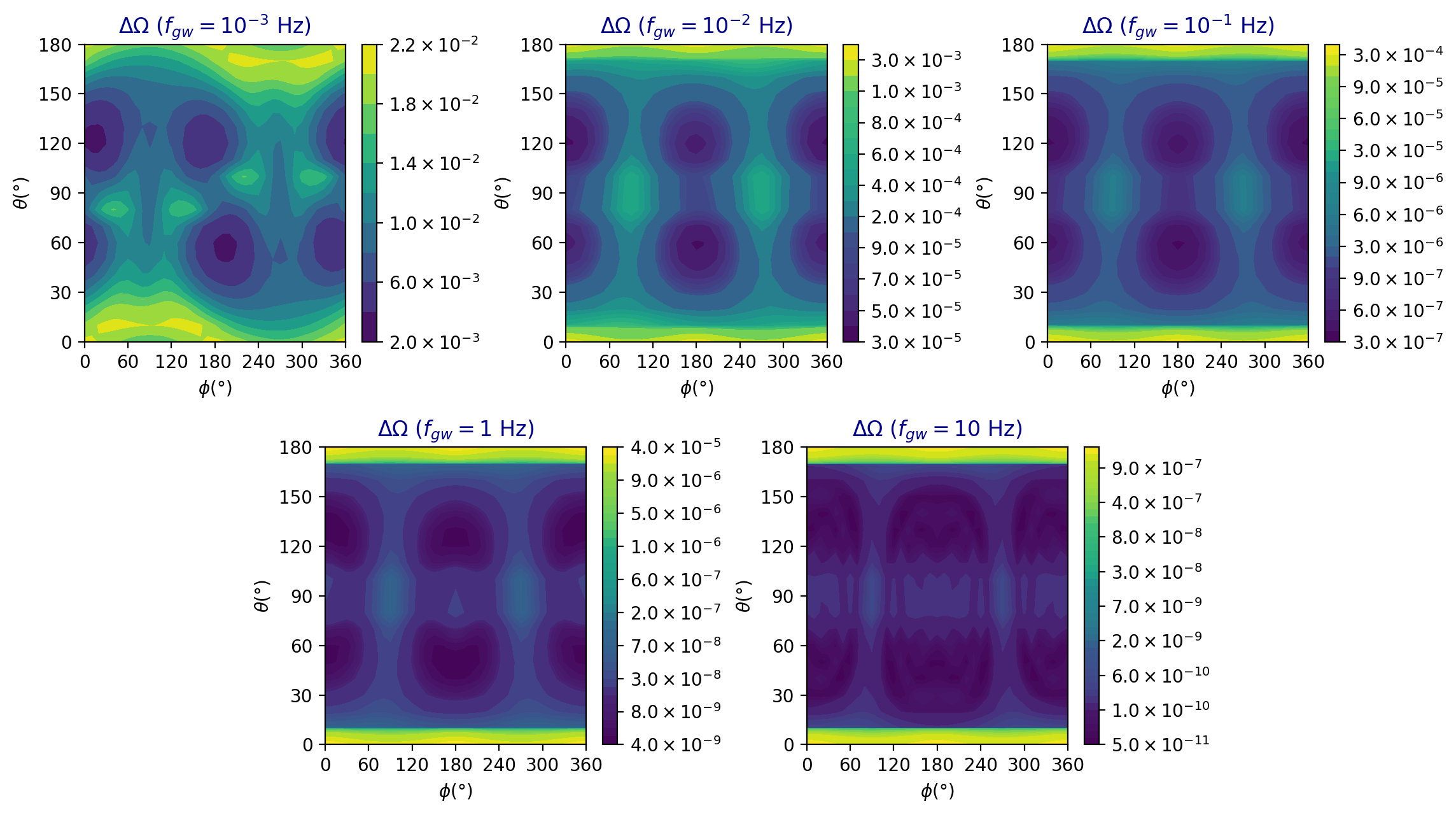}} 
\caption{TianQin angular precision $\Delta \Omega$ as a function of
  the location of the source in the sky and for the selected five~{GW} frequencies. The~average signal-to-noise ratio
  has been taken to be equal to $10$ and the polarization of the waves
  is linear in (\textbf{a}) and circular in (\textbf{b}).}
    \label{fig:TQ_AngR2D}
\end{figure}

In what follows, we present the precisions for different positions of
the source in the sky. Figure~\ref{fig:TQ_AngR2D} shows the TianQin
angular precision $\Delta \Omega$ as a function of the location of the
source, ($\theta, \phi$), for~the following selected GW frequencies:
$f_{gw} = 10^{-3}$, $10^{-2}$, $10^{-1}$, $1$, $10$~Hz. This is done
for (a) linear and (b) circular polarized signals, and~for a
signal-to-noise ratio (averaged over polarization states and source
locations) equal to $10$.

The angular resolutions for both linearly and circularly polarized GWs
show some degradation at $\theta = 0, \pi$ and for some values of the
angle $\phi$, which depend on the frequency of the GW and the polarization state. This is a consequence of the plane of the TianQin array being
almost orthogonal to the plane of the ecliptic. At~$f=10^{-3}$ Hz,
$\theta = 0$, for~instance, and~for linearly polarized waves, we may
notice that at $\phi \simeq \pi/4, 7 \pi/4$, the~angular resolution
degrades by about an order of magnitude w.r.t. its best
value. Similarly, at~$\theta = \pi$, we notice the same degradation at
$\phi = 3 \pi/4, 5 \pi/4$, complementary to the configuration with
$\theta = 0$. We may also observe that the angular precision and its
dynamic range improve throughout the sky as the~{GW}
frequency increases. At~$f_{gw} = 10^{-2}$ Hz, for~example, the~dynamic range in angular resolution for linearly polarized signals is
approximately three orders of magnitude and increases to approximately
four orders of magnitude at $f_{gw} = 10$. For~circularly polarized
signals, the~angular resolution and its dynamic range at
each GW frequency improve further by approximately a factor of ten
over the linear polarization~case.

In Figure~\ref{fig:TQ_EstErr2D_amp}, we turn to the precision in the
reconstructed wave amplitude in terms of the location of the source in
the sky, the~five GW frequencies ($f_{gw} = 10^{-3}$, $10^{-2}$,
$10^{-1}$, $1$, $10$~Hz) and for (a) linearly and (b) circularly polarized
waves, respectively. As~in the case of the angular precision, here the
amplitude precision also shows a loss along the directions
$\theta = 0, \pi$ and for some values of the angle $\phi$. We may also
notice that, independently of the GW polarization, the~
precision in the amplitude increases with the GW frequency in the interval
$[10^{-3} , 10^{-1}]$ Hz and decreases with the GW frequency in the
interval \mbox{$[1 , 10]$ Hz}. This is due to the dependence of the TianQin
sensitivity curve on the GW frequency (see
Figure~\ref{fig:sensitivities}). Also, as~expected, the~precision of the
reconstructed amplitude is better for circularly polarized GW signals,
while the corresponding dynamic ranges over the source sky location
are comparable for the two~polarizations.

\vspace{-12pt}

\begin{figure}[H]
\subfloat[\centering]{\includegraphics[width=0.8\textwidth]{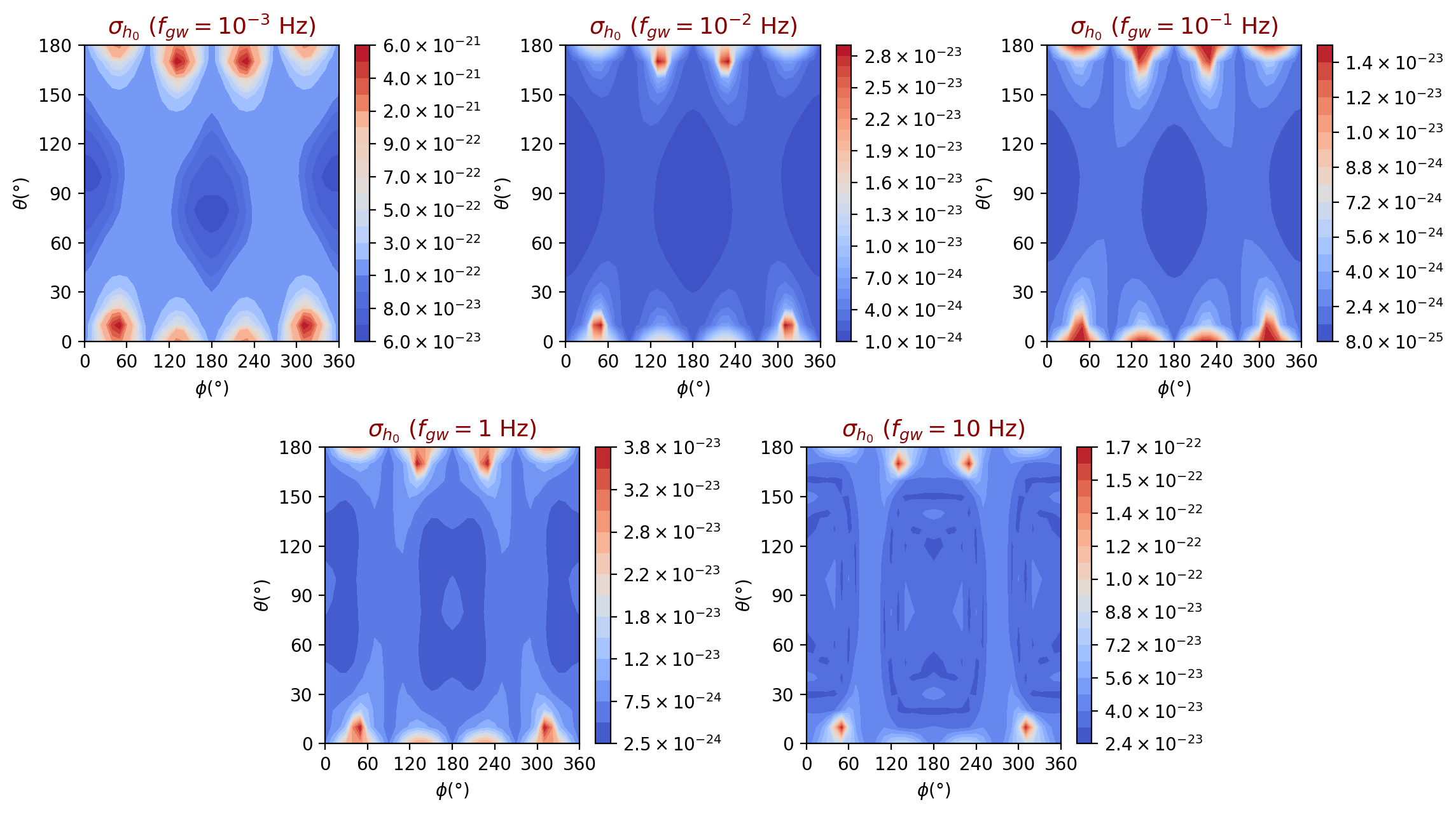}} \\

\caption{\textit{Cont}.}
    \label{fig:TQ_EstErr2D_amp}
\end{figure}

\begin{figure}[H]\ContinuedFloat
\setcounter{subfigure}{1}

\subfloat[\centering]{\includegraphics[width=0.8\textwidth]{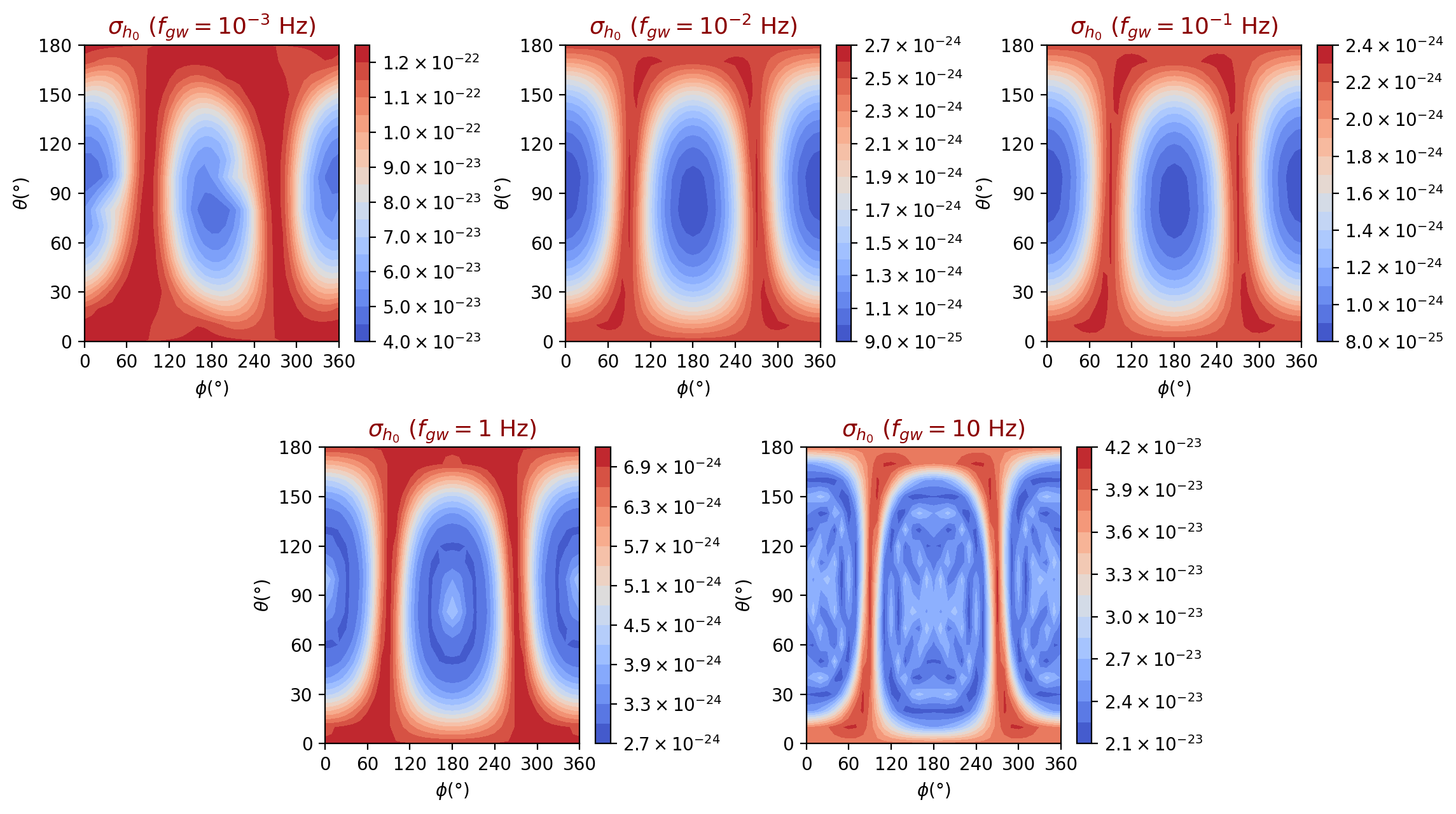}}
    \caption{TianQin~{GW} amplitude precision, $\sigma_{h_0}$, as~      a function of the location of the source in the sky and for
      selected five~{GW} frequencies. The~average
      signal-to-noise ratio has been assumed to be equal to $10$ and
      the polarization of the wave has been chosen to be linear in (\textbf{a})
      and circular in (\textbf{b}).}
    \label{fig:TQ_EstErr2D_amp}
\end{figure}

In the next two sets of contour plots, we finally present the TianQin GW
frequency precisions, $\sigma_{f_{gw}}$, as~functions of
the source sky location, for~the same five GW frequencies considered
in the earlier plots, and~for an average SNR of
$10$. Figure~\ref{fig:TQ_EstErr2D_freq}a shows the precision for
linearly polarized waves, while Figure~\ref{fig:TQ_EstErr2D_freq}b
covers circularly polarized signals. Both contour plots show a dynamic
range equal to approximately $10$ across the entire sky and for all GW
frequencies considered. Although~the difference in magnitude of the
precision between the two polarizations is on average equal to a
factor of $\sqrt{2}$ at the frequencies considered, the~equal-level
contours from the two polarizations show some marked differences in
terms of the location of the source in the sky. Also, like the
previous precision contour plots, here we may notice that at
$\theta = 0, \pi$ and for some values of $\phi$, the~frequency
precision shows some degradation. This is because at these source
locations, the signal-to-noise ratio is penalized by the inclination of
the array w.r.t. the plane of the ecliptic (equal to approximately
$\pi/2$). Optimal precisions are achieved around $\theta = \pi/2$
and $\phi = 0, \pi, 2\pi$ for frequencies in the range
[$10^{-3}, 10^{-1}$] Hz, and~for both linear and circular
polarizations. At~higher frequencies, the~location of the optimal
points changes. This is because at these frequencies, the antenna
patterns become functions of the GW frequency and the
direction-dependent travel time of the GW across the~array.

\begin{figure}[H]
\setcounter{subfigure}{0}
\subfloat[\centering]{\includegraphics[width=0.8\textwidth]{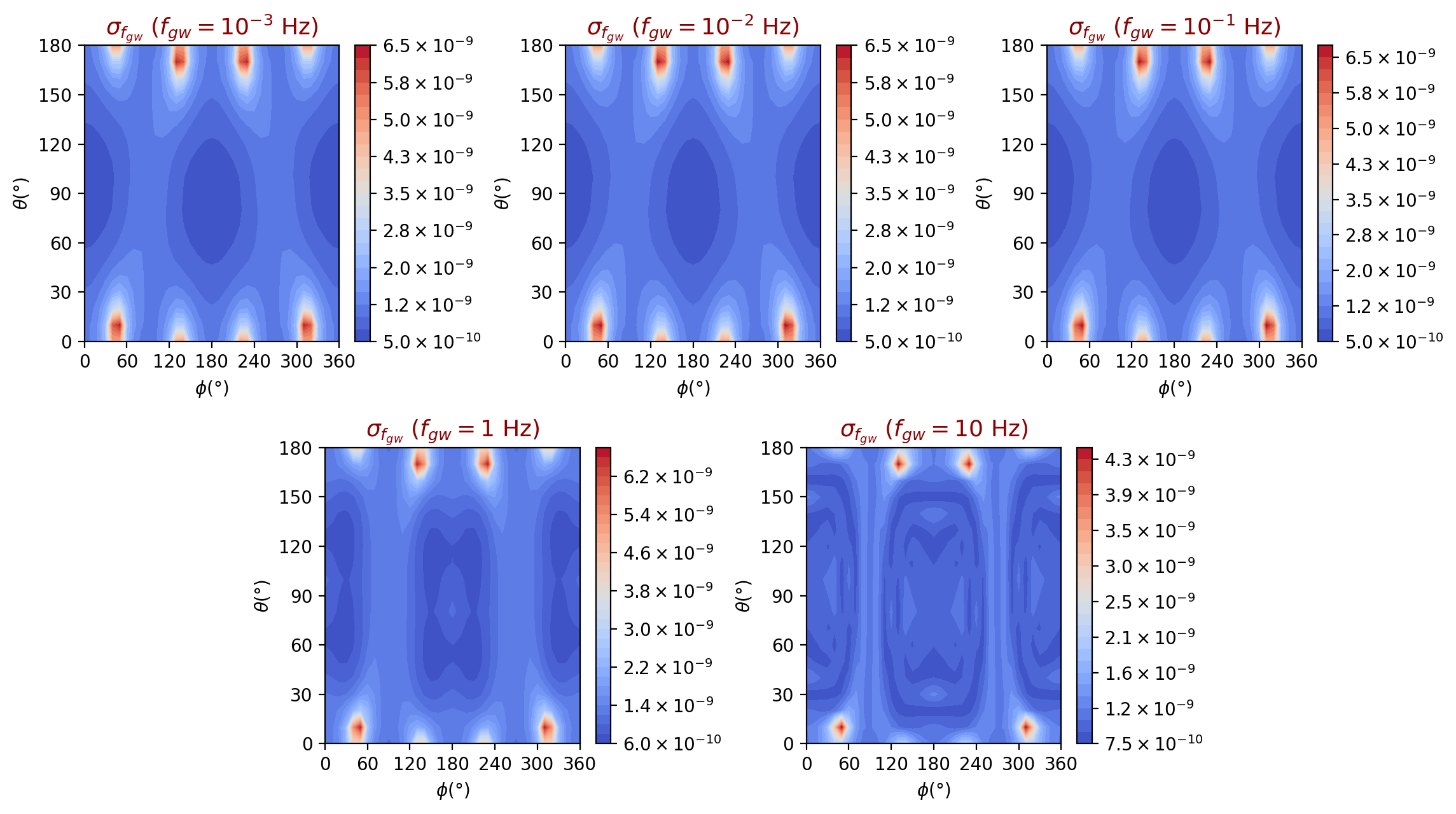}} \\

\caption{\textit{Cont}.}
\label{fig:TQ_EstErr2D_freq}
\end{figure}

\begin{figure}[H]\ContinuedFloat
\setcounter{subfigure}{1}

\subfloat[\centering]{\includegraphics[width=0.8\textwidth]{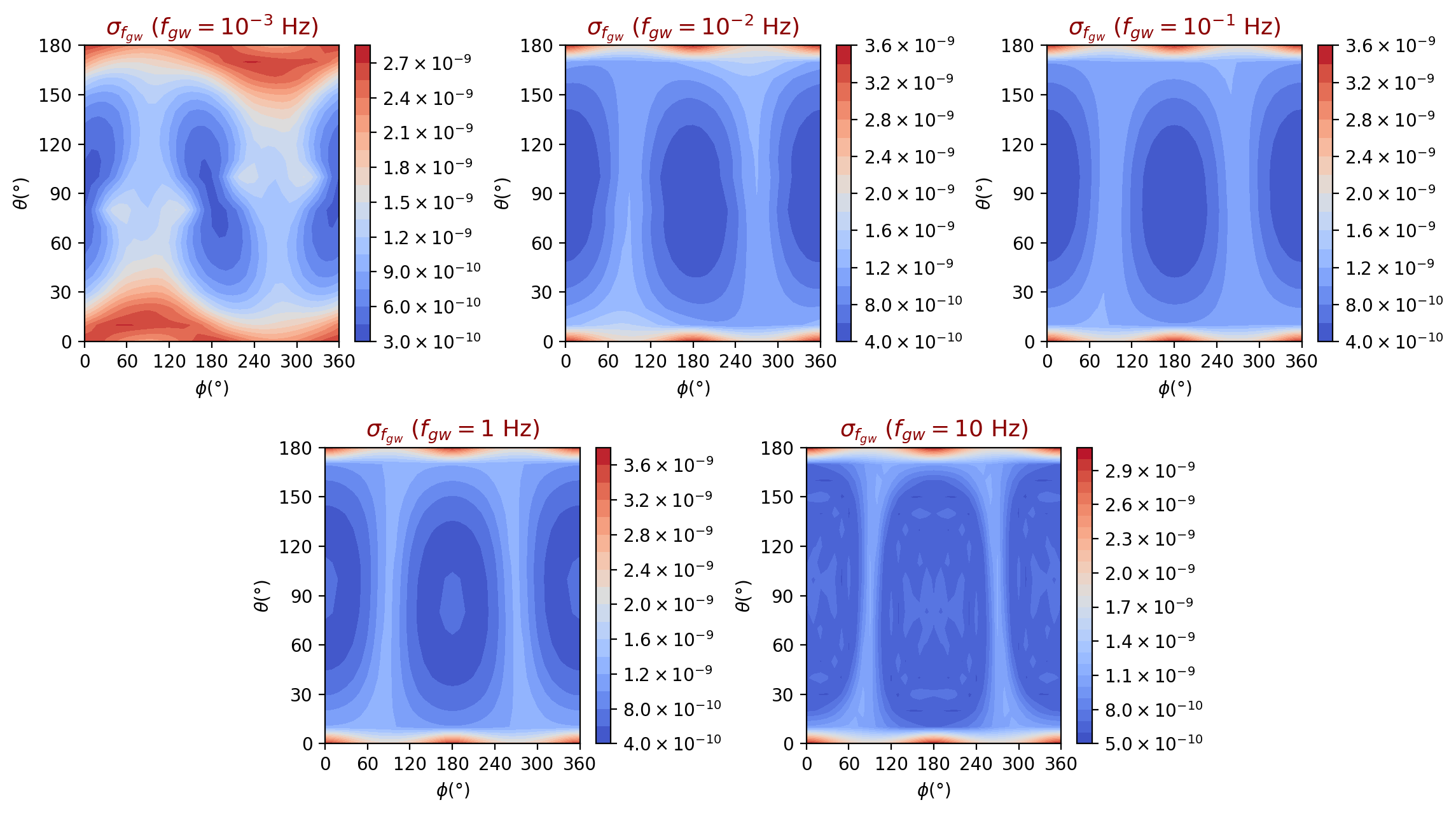}}
\caption{TianQin estimated precision of the~{GW} frequency,
  $\sigma_{f_{gw}}$, as~a function of the location of the source in
  the sky and for selected five~{GW}
  frequencies. The~average signal-to-noise ratio has been assumed
  to be equal to $10$ and the polarization of the wave has been
  chosen to be linear in (\textbf{a}) and circular in (\textbf{b}).}
\label{fig:TQ_EstErr2D_freq}
\end{figure}

\subsection{\texorpdfstring{gLISA Parameter Estimation~Errors}{gLISA Parameter Estimation Errors}}

\begin{figure}[H]
\includegraphics[width=0.8\textwidth, angle=0.0]{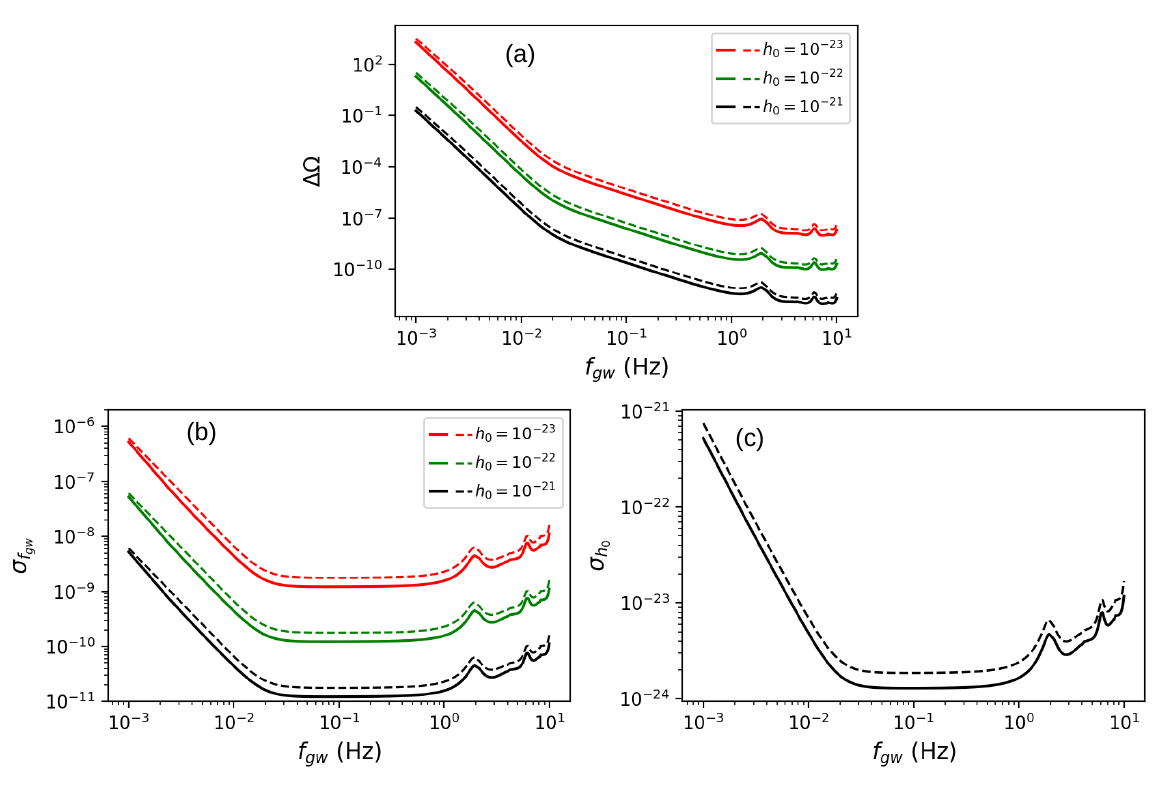}
\caption{The gLISA angular $\Delta \Omega$ (\textbf{a}), frequency
  $\sigma_{f_{gw}}$ (\textbf{b}), and amplitude $\sigma_{h_0}$ (\textbf{c}) precisions as
  functions of the~{GW} frequency, $f_{gw}$, for~the
  same source location and the three values of the~{GW}
  amplitude considered in the previous subsection for
  TianQin. Continuous lines are for circular polarization, while
  dashed lines are for linear polarization. See text for a detailed
  discussion. {As in the case of the TianQin mission, the~angular precision is not proportional to the sensitivity curve as it includes an additional frequency dependence captured by Equation~(\ref{DeltaOmega}).}}
\label{GL_AngRes_VS_fre}
\end{figure}

The analysis for gLISA follows a similar approach to that described
earlier for TianQin, with~key differences arising from its distinct
orbit and design. In~Figure~\ref{GL_AngRes_VS_fre}a, we present the
angular resolution, $\Delta \Omega$, as~a function of $f_{gw}$ for the
same source location selected for TianQin and the same three GW
amplitudes ($h_0=10^{-23}$, $h_0=10^{-22}$, $h_0=10^{-21}$). The~results are shown for circular (solid lines) and linear (dashed lines)
polarizations, which differ by approximately a factor of two as
well. Compared to TianQin, the~angular error for gLISA is larger at
lower frequencies; above $f_{gw} \approx 6 \times 10^{-1}$\text{Hz},
the performance of both detectors becomes comparable in terms of the
source location error. A~similar trend is observed in the precision of
the frequency and amplitude, as~shown in Figure~\ref{GL_AngRes_VS_fre}b,c, respectively. Both precisions follow the behavior of the
sensitivity curve, presented in Figure~\ref{fig:sensitivities}, with~lower precision at low frequencies, better precision in the
intermediate range, and~degraded precision above $1~\text{Hz}$.

The angular resolution of gLISA, $\Delta \Omega$, as~a function of the
source location in the sky ($\theta, \phi$), is presented in
Figure~\ref{fig:GL_AngR2D} for linear (a) and circular (b)
polarizations, and~for the same~{GW} frequencies
selected for TianQin. A~degradation in angular resolution is observed
around $\theta = \pi/2$, which becomes less pronounced at higher
frequencies due to the increased sensitivity of the detector in this
part of the accessible band and changes with the polarization state of
the wave. At~$f = 10^{-3}$ Hz and $\theta = \pi/2$, the~angular
resolution of linearly polarized waves degrades by approximately an
order of magnitude compared to its best value. In~contrast, at~$f = 1$
Hz, the~localization error spans three orders of magnitude between its
minimum and maximum~values.

The differences in orders of magnitude between the minimum and maximum
errors show no significant variation when comparing linear and
circular polarizations. Moreover, in~all cases, the~best angular
resolutions are achieved for the source locations around $\phi = 0$,
$\pi$, and~$2\pi$, and~$\theta \approx$ $2\pi/9$, $7\pi/9$, indicating
optimal sensitivity in these~regions.

It should be noted that the region of lower precision around
$\theta = \pi/2$ is ``complementary'' to the regions of lower precision
estimated for TianQin. This distinction arises from the differences in
orbital configurations: TianQin's orbital plane is nearly
perpendicular to the ecliptic, whereas gLISA operates with a
1.5$^{\circ}$ tilt relative to the celestial equator. Another aspect
to note is that the maximum angular error for gLISA is seven times lower
than that of TianQin at a frequency of $1$~Hz and for circularly
polarized~waves.

\begin{figure}[H]
\setcounter{subfigure}{0}
\subfloat[\centering]{\includegraphics[width=0.8\textwidth]{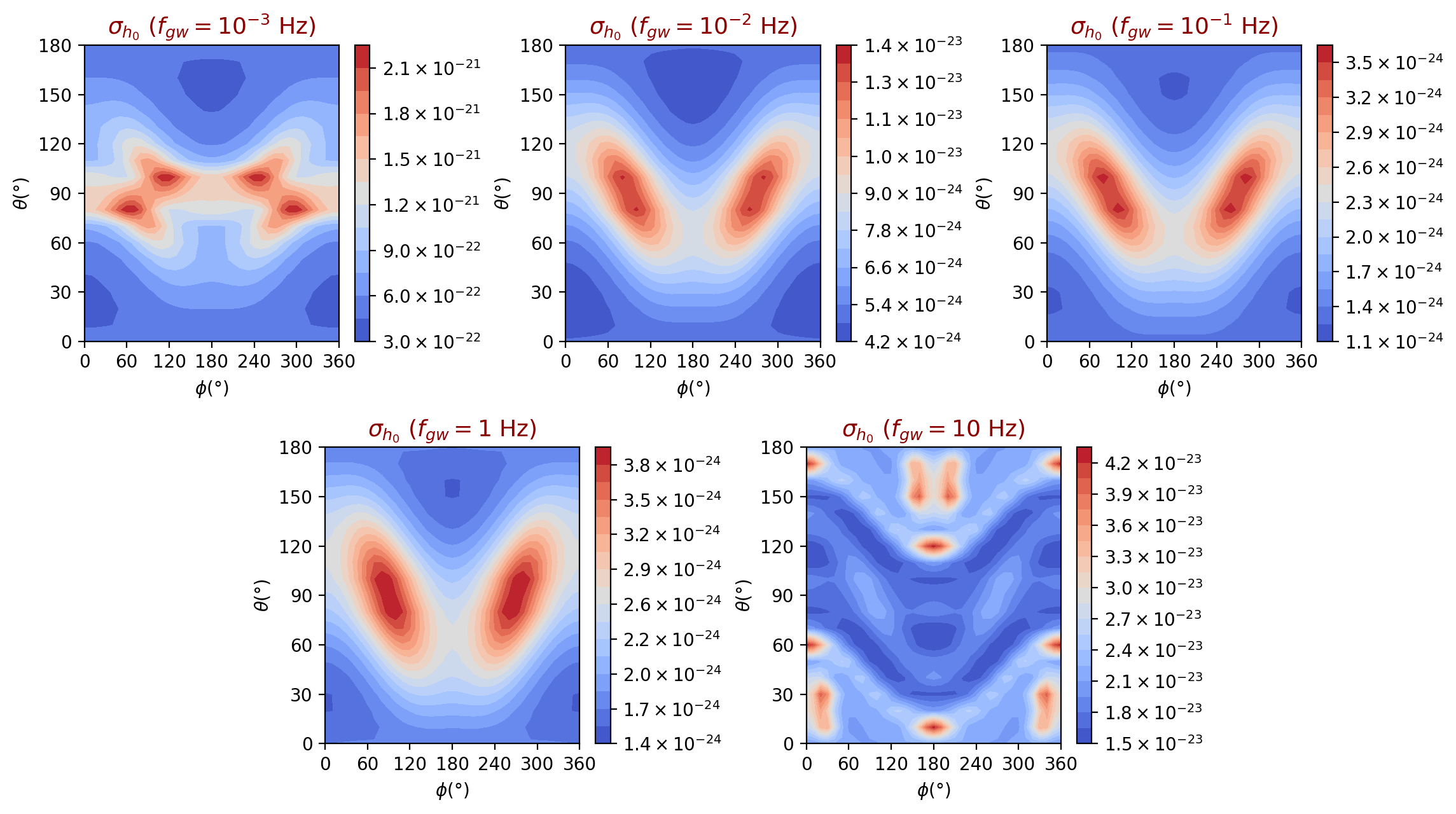}} \\
\subfloat[\centering]{\includegraphics[width=0.8\textwidth]{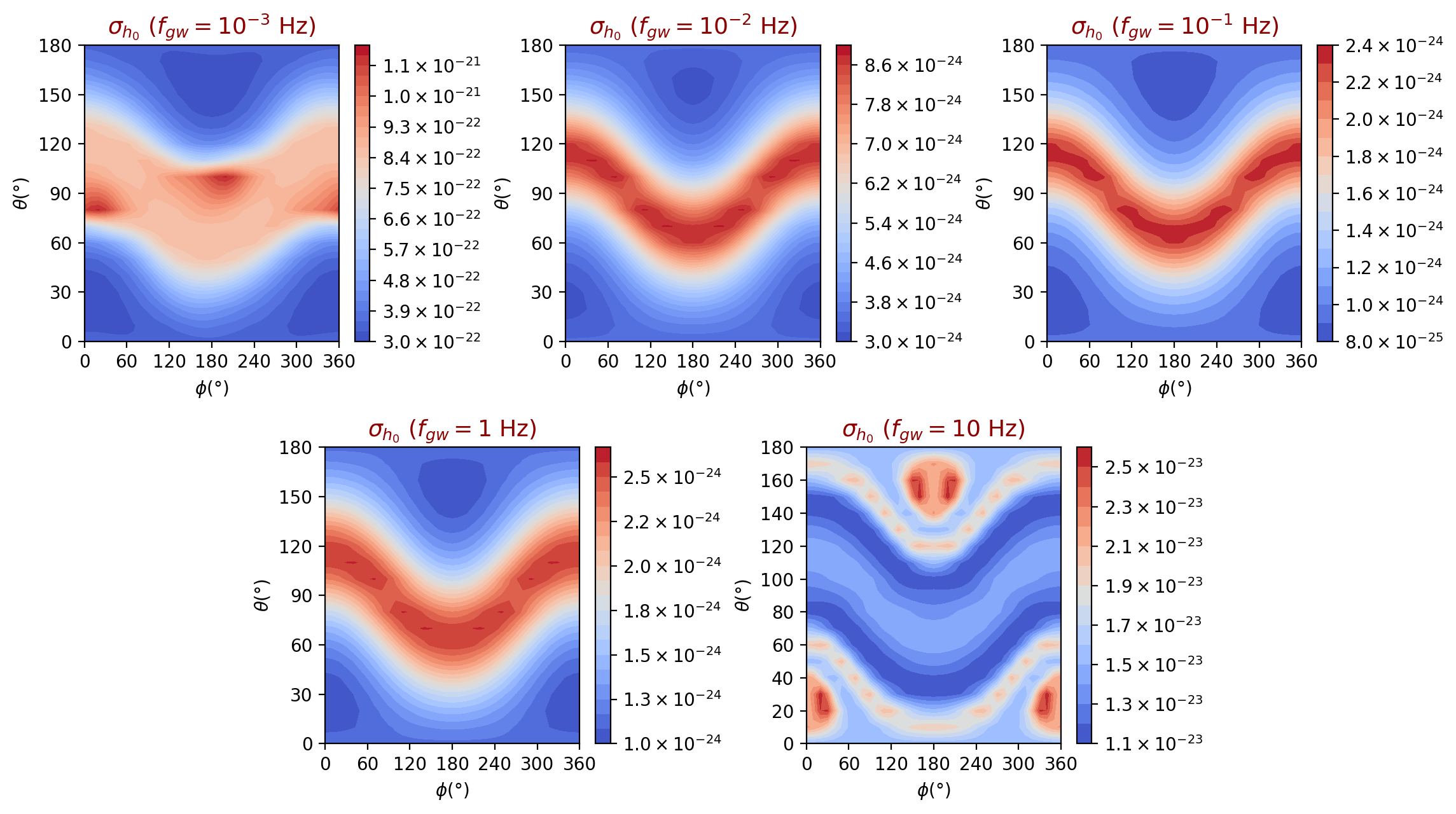}}
\caption{gLISA~{GW} amplitude precision $\sigma_{h_0}$ as
      a function of the location of the source in the sky and for
      five selected~{GW} frequencies. The~average
      signal-to-noise ratio has been assumed to be equal to $10$ and
      the polarization of the wave has been chosen to be linear in (\textbf{a})
      and circular in (\textbf{b}).}
    \label{fig:GL_EstErr2D_amp}
\end{figure}

In Figure~\ref{fig:GL_EstErr2D_amp}, we show the precisions in the~{GW} amplitude as functions of the GW frequency, the~location of the source in the sky, and for linearly (a) and circularly
(b) polarized signals. It can be seen that the minimum and maximum
values differ by only a factor of approximately three, indicating
relatively small variation between different frequencies. Similarly to
angular precision, these graphs show a loss around $\theta = \pi/2$
and its proximity, which varies with the $\phi$ angle. In~addition, as~for TianQin, the~amplitude precisions are slightly better for circular
than for linear~polarization.

\begin{figure}[H]
\subfloat[\centering]{\includegraphics[width=0.8\textwidth]{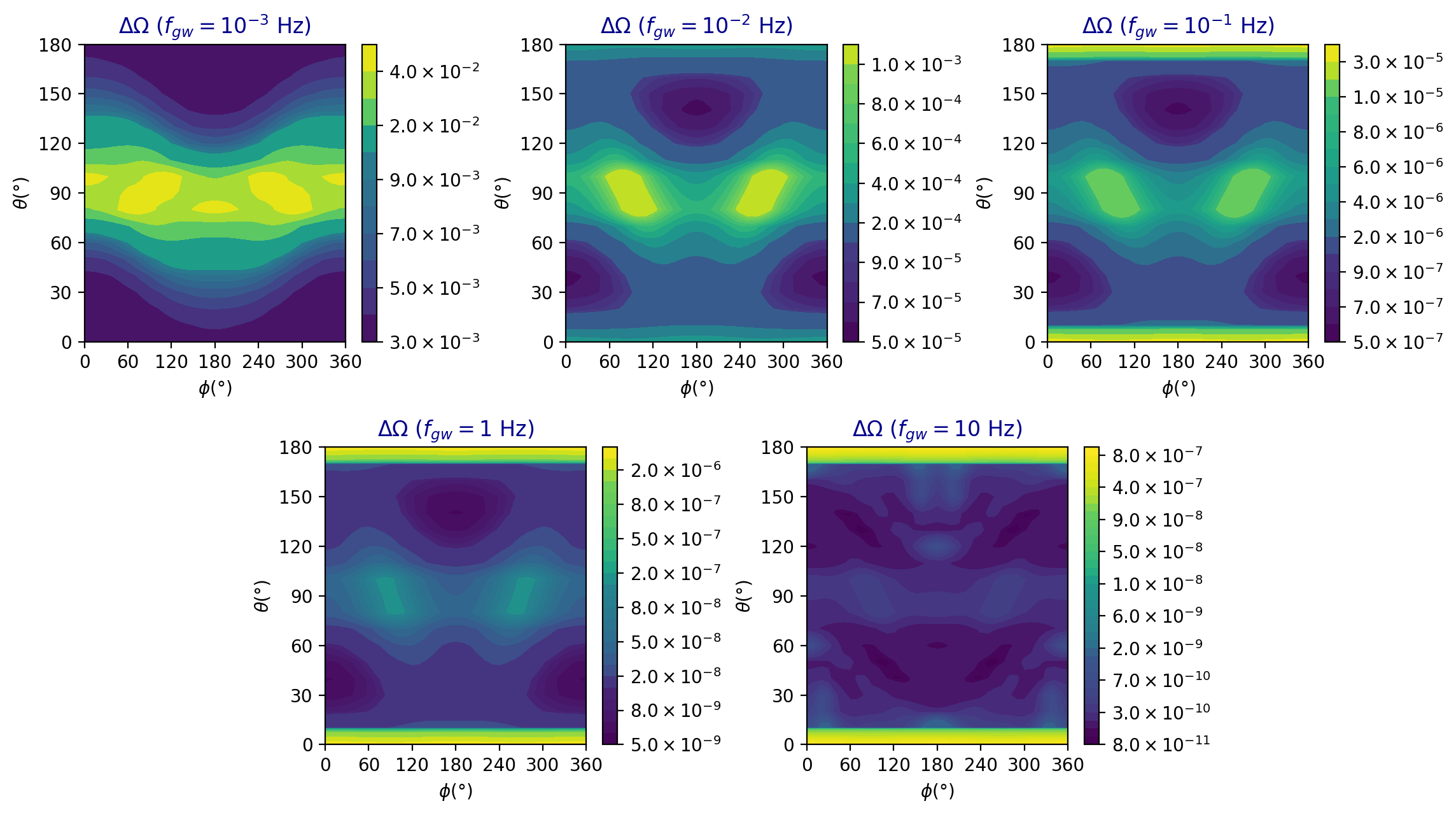}} \\\caption{\textit{Cont}.}
    \label{fig:GL_AngR2D}
\end{figure}

\begin{figure}[H]\ContinuedFloat
\setcounter{subfigure}{1}
\subfloat[\centering]{\includegraphics[width=0.8\textwidth]{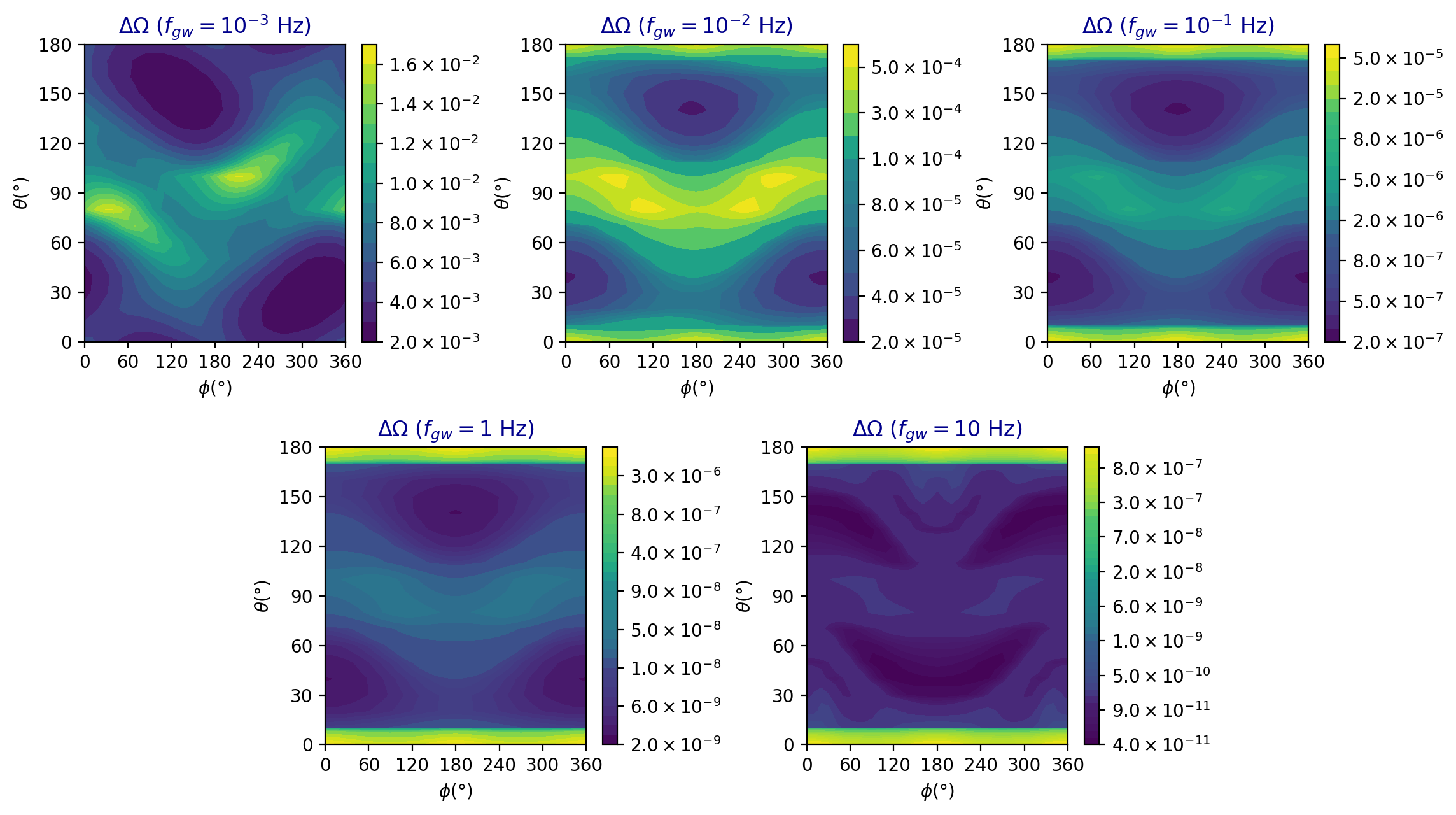}}
\caption{gLISA angular precision $\Delta \Omega$ as a function of the
      location of the source in the sky and for the selected five~{GW} frequencies. The~average signal-to-noise
      ratio has been taken to be equal to $10$ and the polarization of
      the waves is linear in (\textbf{a}) and circular in (\textbf{b}).}
    \label{fig:GL_AngR2D}
\end{figure}

\vspace{-12pt}

\begin{figure}[H]
\subfloat[\centering]{\includegraphics[width=0.8\textwidth]{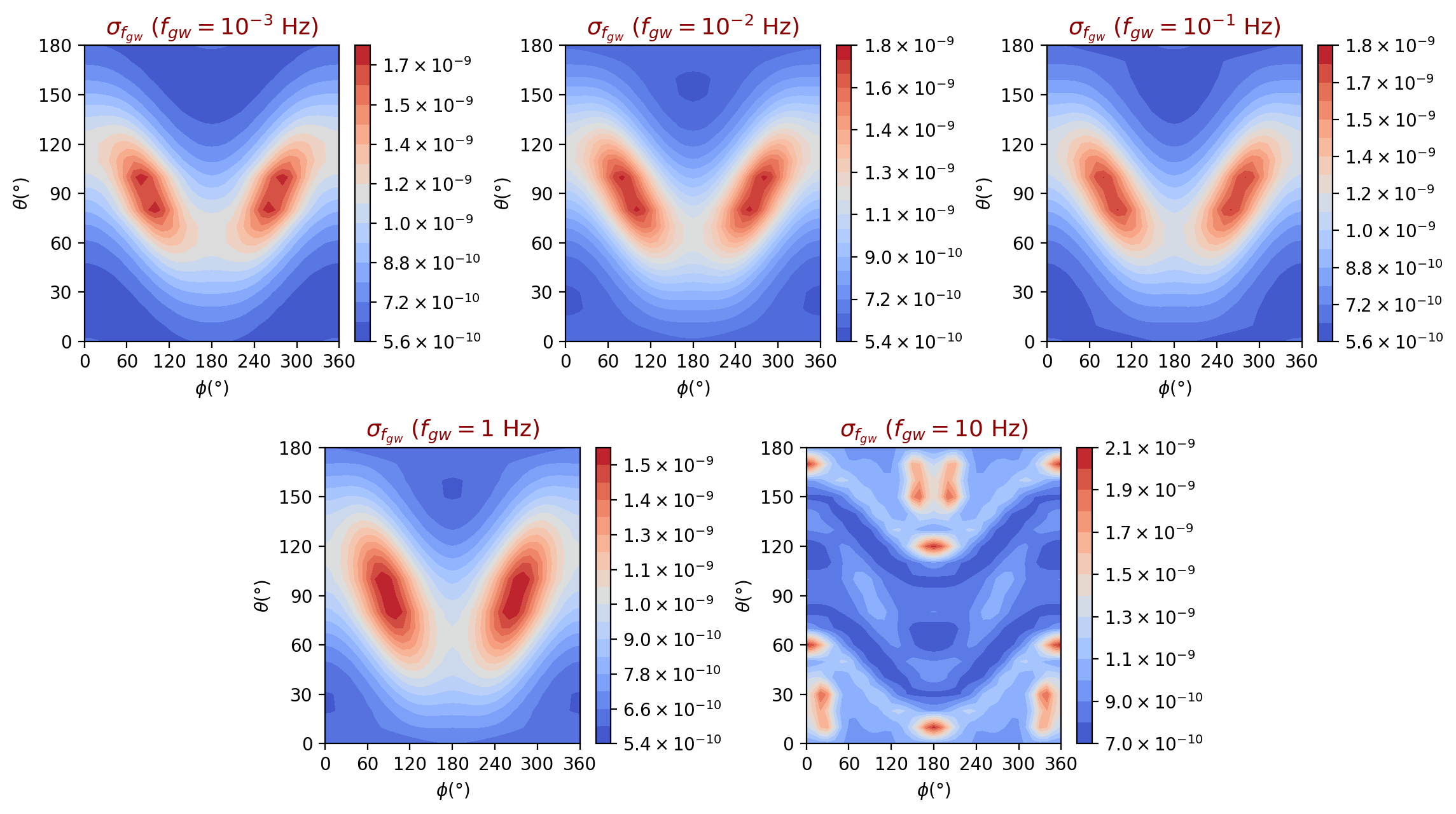}} \\
\subfloat[\centering]{\includegraphics[width=0.8\textwidth]{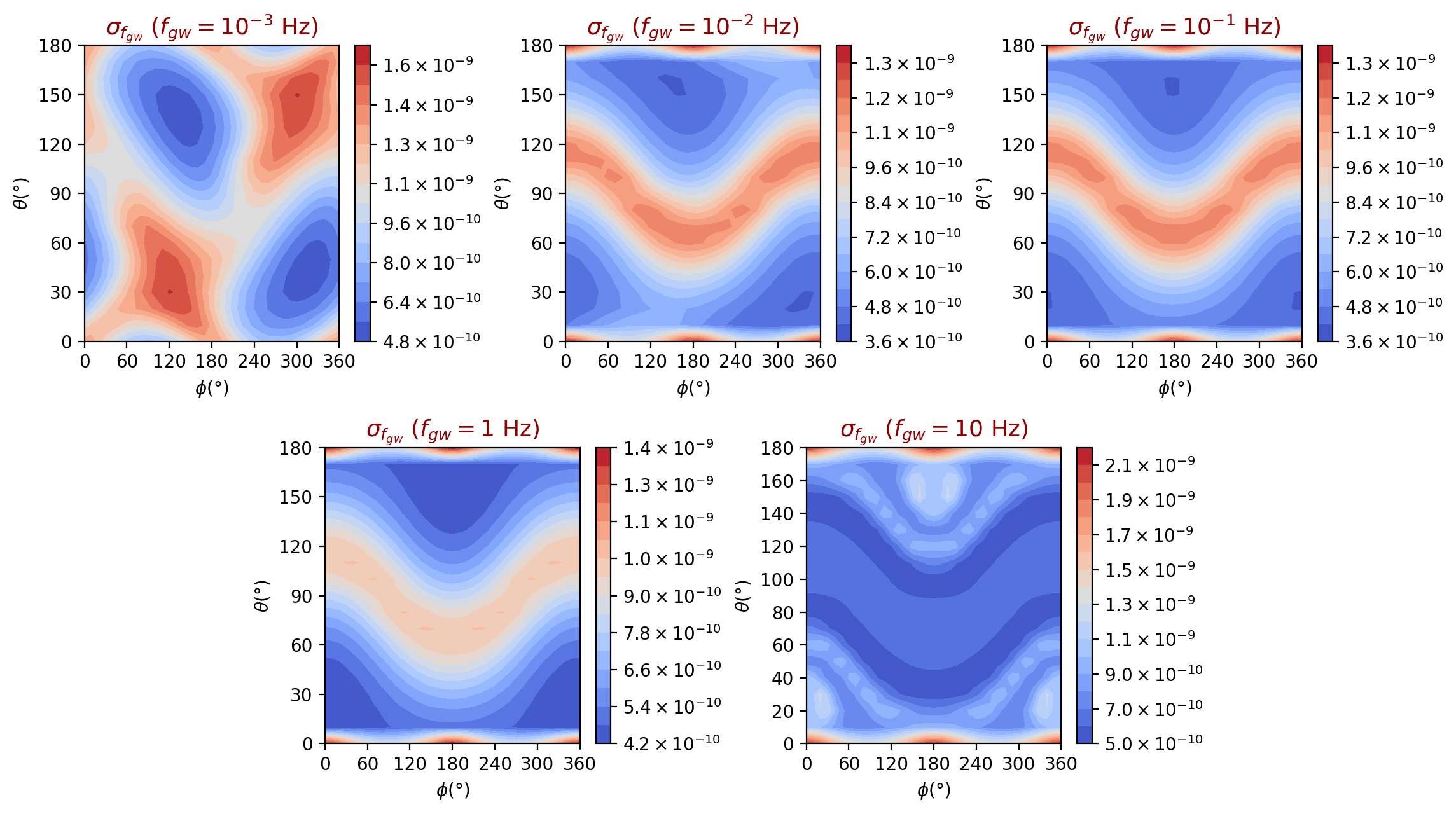}}
\caption{gLISA~{GW} frequency estimated precision
      $\sigma_{f_{gw}}$ as a function of the location of the source in
      the sky and for selected five~{GW}
      frequencies. The~average signal-to-noise ratio has been assumed
      to be equal to $10$ and the polarization of the wave has been
      chosen to be linear in (\textbf{a}) and circular in (\textbf{b}).}
    \label{fig:GL_EstErr2D_freq}
\end{figure}

The errors in the reconstructed~{GW} frequency,
$f_{gw}$, as~a function of the source's location, are presented in
Figure~\ref{fig:GL_EstErr2D_freq}. They exhibit a structure very
similar to that observed in the graphs for the amplitude, except~for
that at frequency $10^{-3}~\text{Hz}$. For~linear polarization, the~error at $10^{-3}$ Hz follows the same behavior as the other
frequencies, while for circular polarization, the~optimal responses
are now located at $(\theta, \phi) = (\pi/4, 8\pi/9) $ and
\mbox{$ (\theta, \phi) = (7\pi/9, 7\pi/9) $}. Additionally, the~error dynamic
range for circular and linear polarizations is nearly~identical.
\vspace{-3pt}

\subsection{\texorpdfstring{gLISA$_d$}{gLISAd} Parameter Estimation~Errors}

\begin{figure}[H]
\includegraphics[width=0.8\textwidth, angle=0.0]{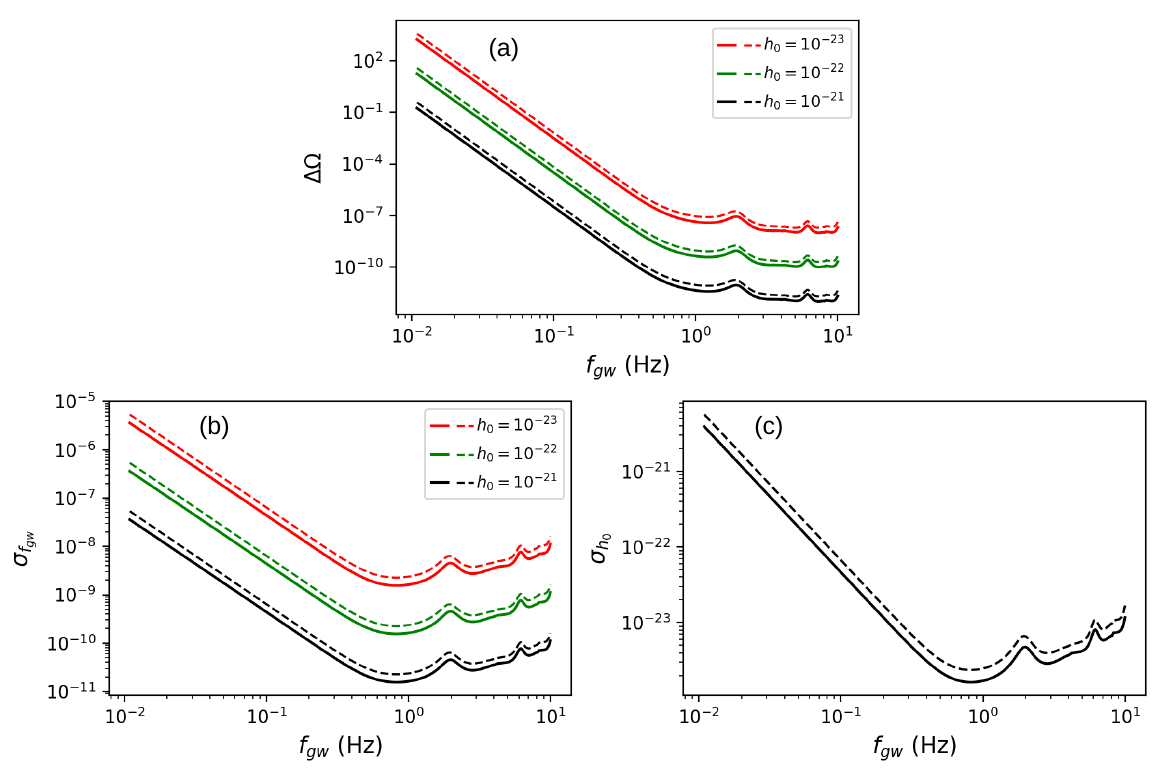}
\caption{The angular $\Delta \Omega$ (\textbf{a}), frequency $\sigma_{f_{gw}}$
  (\textbf{b}), and amplitude $\sigma_{h_0}$ (\textbf{c}) precisions as functions of the~{GW} frequency, $f_{gw}$, and~for three values of the~{GW} amplitude. The~angular location of the GW signal
  has been selected as in the previous two corresponding sets of
  graphs for TianQIn and gLISA. Continuous lines are for circular
  polarization, while dashed lines are for linear polarization. See
  the text for a detailed discussion. {Again, the~angular precision is not proportional to the sensitivity curve as it includes an additional frequency dependence captured by Equation~(\ref{DeltaOmega}).}}
\label{IM_AngRes_VS_fre}
\end{figure}

The performance of gLISA$_d$ at low frequencies, shown in
Figure~\ref{IM_AngRes_VS_fre}a--c, reflects its low sensitivity in
this part of the frequency band. However, at~higher frequencies,
gLISA$_d$ achieves precisions comparable to those of gLISA and
slightly better than those characterizing TianQin. Part (a) presents
the angular precision as a function of the GW frequency for the same
source location considered earlier for TianQin and gLISA. The~results
for the precision at $f = 10^{-3}$~Hz are not presented due to the
poorer sensitivity of this detector at that frequency. At~$f = 10^{-2}$~Hz, the~angular error is six orders of magnitude worse
than that of TianQin and five orders of magnitude worse than that of
gLISA. However, at~\mbox{$f_{gw} = 1$ Hz} and above, the~gLISA$_d$ precisions
in angular, amplitude, and frequency reconstructions become equal to
those of gLISA and better than those of TianQin. In~part (b), the~frequency error closely follows the sensitivity curve for the three GW
amplitudes considered. Finally, part (c) illustrates the amplitude
error, which is independent of the signal amplitude and is therefore
represented using a single~color.

In the following contour plots, the~precision of each parameter is now
presented not for a specific source, but~as a function of the angular
positions ($\theta$ and $\phi$) at frequencies ranging from $10^{-2}$
to $10$ Hz.

\begin{figure}[H]
\subfloat[\centering]{\includegraphics[width=0.8\textwidth]{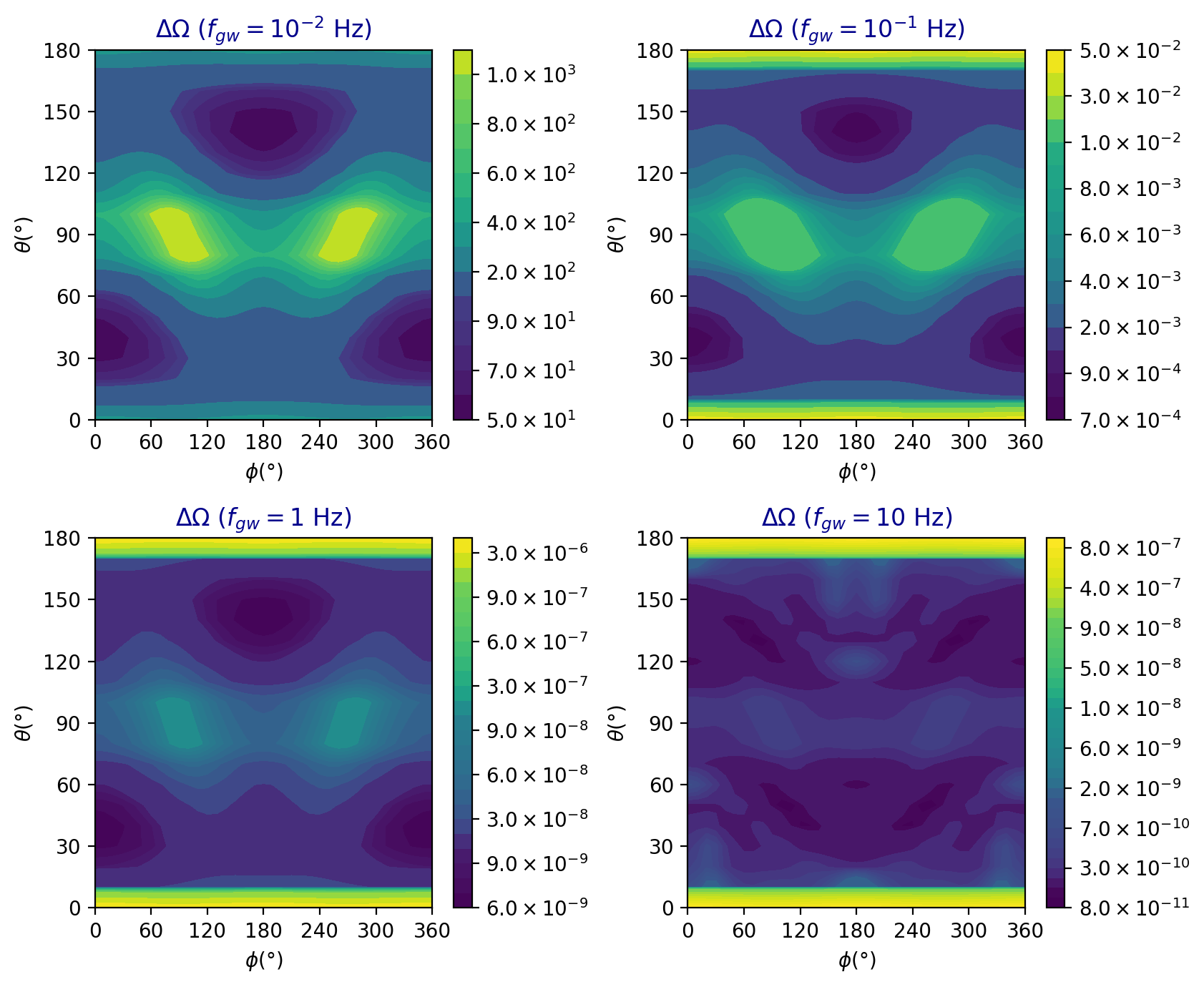}} \\
\subfloat[\centering]{\includegraphics[width=0.8\textwidth]{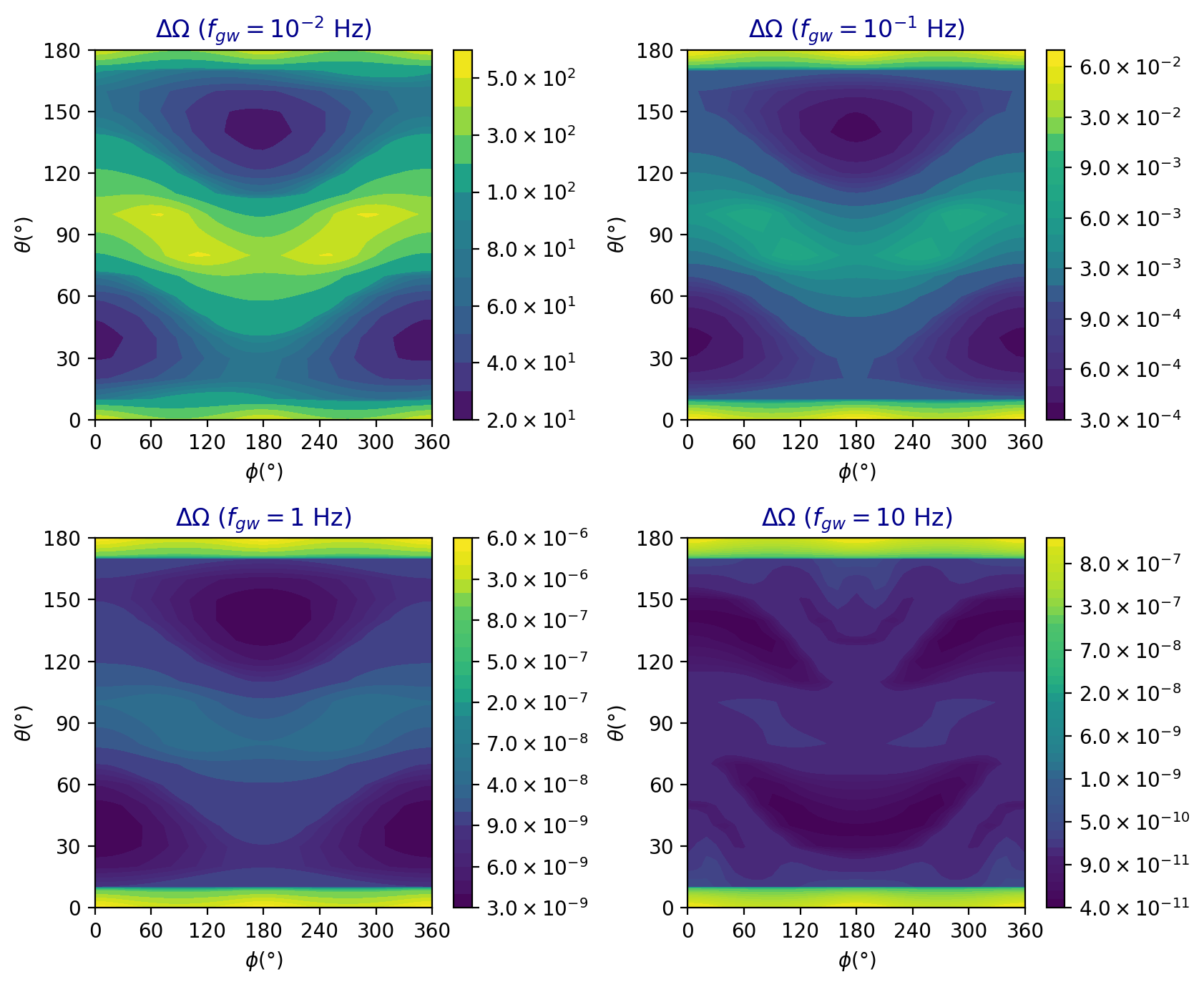}}
\caption{gLISA$_d$ angular precision $\Delta \Omega$ as a function of
  the location of the source in the sky and for the four selected ~{GW} frequencies. The~{GW} amplitudes
  are those given in Table~\ref{tab:snr_values}, which result in an
  average SNR of 10 in gLISA. The~polarization of the waves is linear
  in (\textbf{a}) and circular in (\textbf{b}).}
    \label{fig:IM_AngR2D}
\end{figure}

Figure~\ref{fig:IM_AngR2D} shows the angular precision for
linearly polarized waves (a) and circularly polarized waves (b). As~expected, due to the detector geometry and trajectory, the~contour
plot topologies are identical to those of gLISA, with~maximum and
minimum error values larger due to the poorer sensitivity at
frequencies smaller than $1$ Hz. However, at~frequencies larger than
$1$ Hz, we recover the same contour lines shown for gLISA as at these
frequencies gLISA and gLISA$_d$ achieve the same SNR. A~similar
behavior is also evident in the amplitude and frequency precision
contours discussed~below.

\begin{figure}[H]
\subfloat[\centering]{\includegraphics[width=0.8\textwidth]{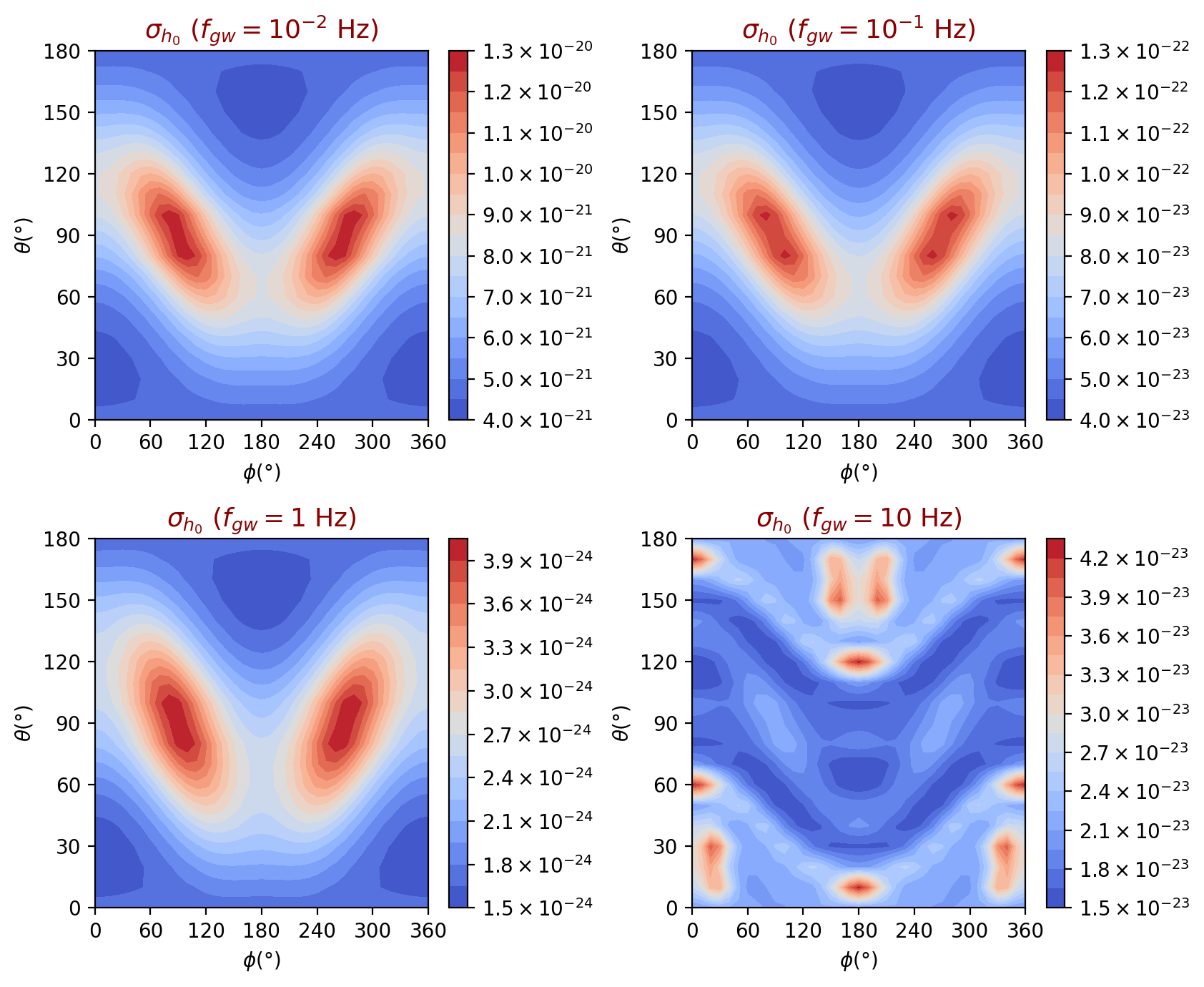}} \\
\subfloat[\centering]{\includegraphics[width=0.8\textwidth]{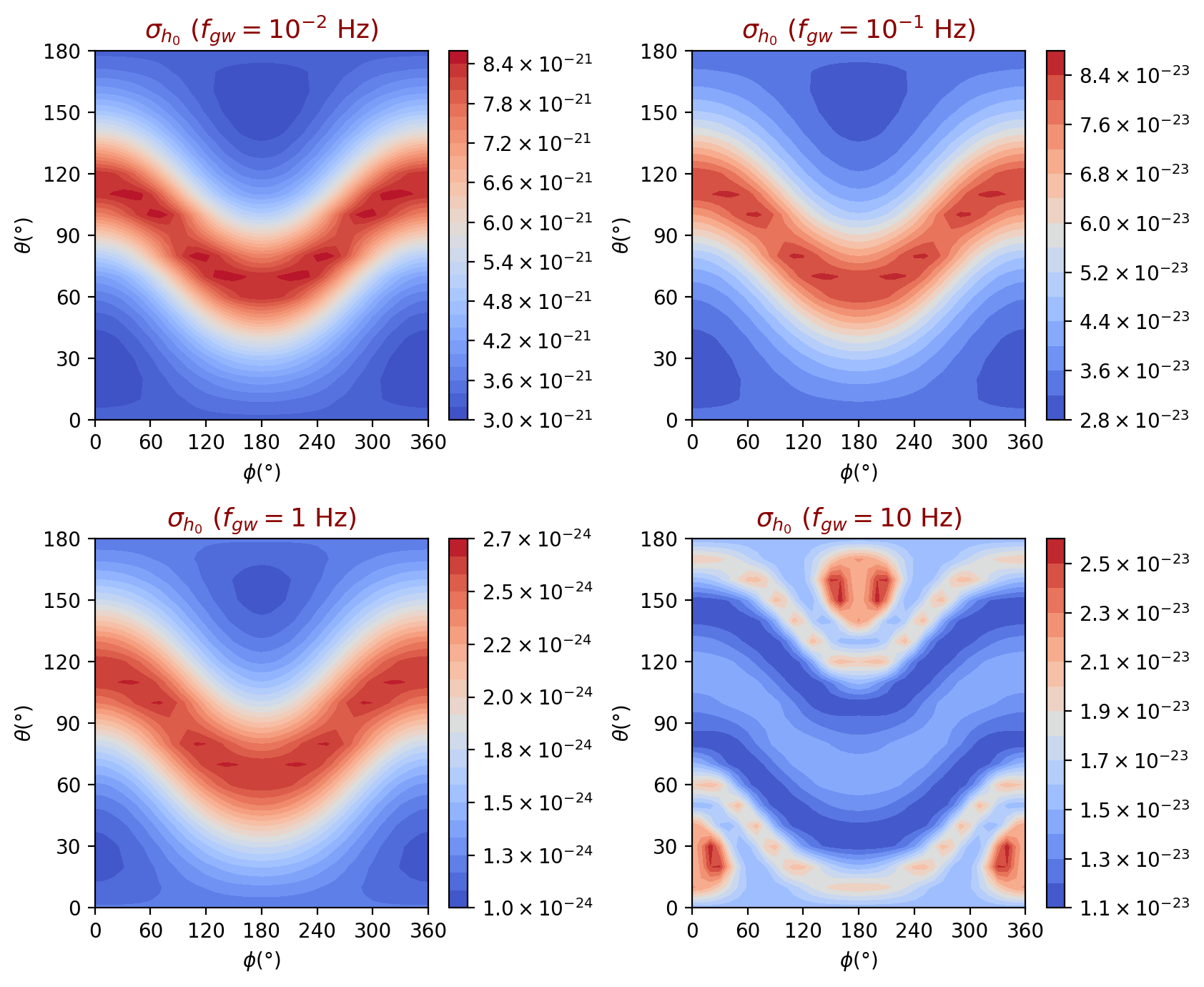}}
\caption{gLISA$_d$~{GW} amplitude precision
  $\sigma_{h_0}$ as a function of the location of the source in the
  sky and for four selected {GW} frequencies. The~{GW} amplitudes are those given in
Table~\ref{tab:snr_values}, which result in an average SNR of 10 in
  gLISA. The~polarization of the waves is linear in (\textbf{a}) and circular
  in (\textbf{b}).}
    \label{fig:IM_EstErr2D_amp}
\end{figure}

\begin{figure}[H]
\subfloat[\centering]{\includegraphics[width=0.8\textwidth]{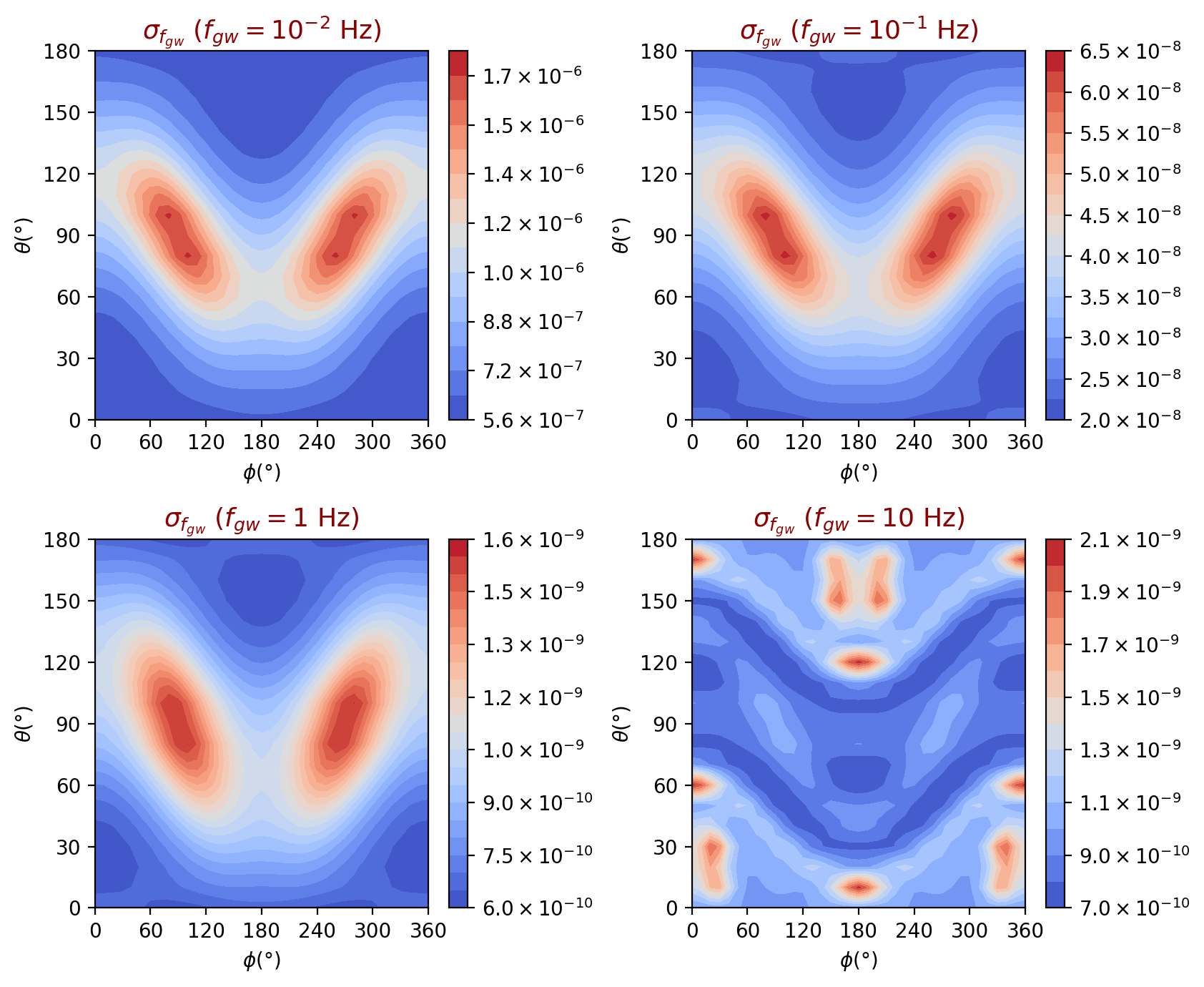}} \\
\subfloat[\centering]{\includegraphics[width=0.8\textwidth]{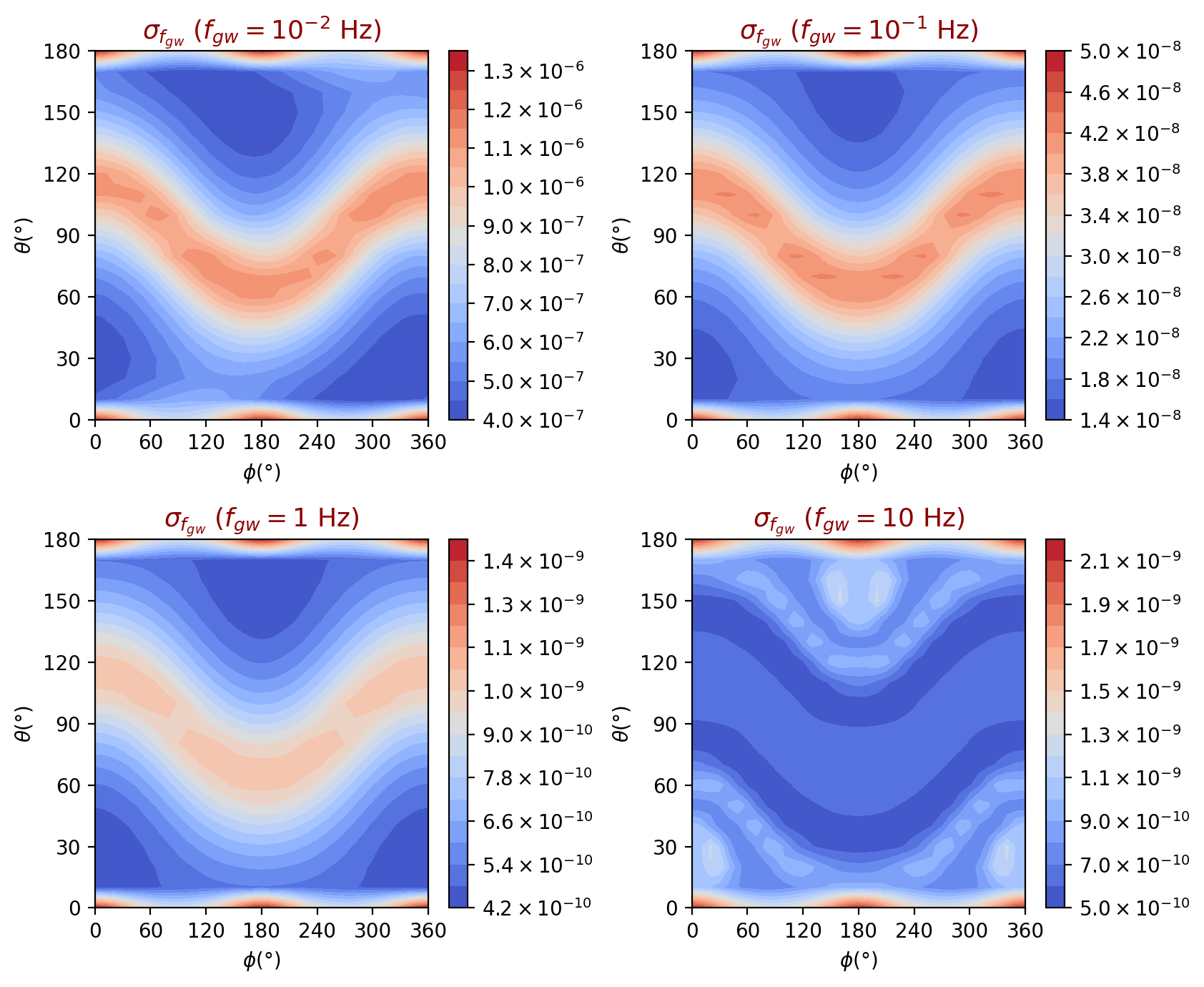}}
\caption{gLISA$_d$~{GW} frequency estimated precision
  $\sigma_{f_{gw}}$ as a function of the location of the source in the
  sky and for selected four~{GW} frequencies. The~{GW} amplitudes are those given in
Table~\ref{tab:snr_values}, which result in an average SNR of 10 in
  gLISA. The~polarization of the waves is linear in (\textbf{a}) and circular
  in (\textbf{b}).}
    \label{fig:IM_EstErr2D_freq}
\end{figure}

The amplitude precisions are given in
Figure~\ref{fig:IM_EstErr2D_amp}, for~linearly polarized (a) and
circularly polarized signals, while the frequency precisions are shown
in Figure~\ref{fig:IM_EstErr2D_freq}a,b.

It is interesting to estimate the~{GW} amplitudes, at~the five selected frequencies, which would make gLISA$_d$ achieve an
average SNR of 10 like gLISA. From~Table~\ref{tab:snr_values}, it is
easy to infer that, at~$10^{-3}$ Hz, an~amplitude
$h_0 = 6.86 \times 10^{-18}$ would give an SNR of 10 in
gLISA$_d$. Instead, at~$10^{-2}$ Hz, an~amplitude
$h_0 = 6.86 \times 10^{-20}$ would be required, while at $10^{-1}$ Hz,
a smaller amplitude of $h_0 = 6.85 \times 10^{-22}$ would be
needed. These signals could be emitted by either super-massive black
holes in the lower part of the band or stellar-mass binary black-holes
at higher frequencies. As~mentioned above, at~frequencies above $1$ Hz
gLISA$_d$ achieves the same sensitivity as gLISA, matching its
precisions in this frequency~band.

\unskip

\section{Summary of the Results and~Conclusions}
\label{SecV}

We presented a Fisher information matrix study of the parameter
estimation precisions achievable with a class of future space-based,
mid-band, {GW} interferometers observing monochromatic
signals. {Mid-band gravitational wave detectors have the potential to play an important role in enabling multi-band observations when operated in conjunction with longer arm length interferometers (such as LISA and Taiji) and ground-based interferometers. By~bridging the frequency gap between these instruments, mid-band detectors can extend the observational window allowing the tracking of GW sources across a wide frequency range---see, e.g., \cite{Sesana,tinto2016coherent}---in~principle from $10^{-4}$ to $10^{3}$~Hz.}

{In this work}, we analyzed monochromatic signals observed by the TianQin
mission, gLISA (a LISA-like interferometer in a geosynchronous orbit),
and a de-scoped gLISA mission, gLISA$_d$, characterized by an
acceleration noise level that is three orders of magnitude worse than
that of gLISA. We found that all three missions achieve their best
angular source reconstruction precision in the higher part of their
accessible frequency band, with~an error box better than $10^{-10}$~sr
in the frequency band [$10^{-1}, 10$]~Hz when observing a
monochromatic~{GW} signal of amplitude $h_0 = 10^{-21}$
and incoming from a given direction. In~terms of their reconstructed
frequencies and amplitudes, TianQin achieves its best precisions in
both quantities in the frequency band [$10^{-2}, 4 \times 10^{-1}$]
Hz, with~a frequency precision $\sigma_{f_{gw}} = 2 \times 10^{-11}$
Hz and an amplitude precision $\sigma_{h_0} = 2 \times
10^{-24}$. gLISA matches these precisions in a frequency band slightly
higher than that of TianQin, [$3 \times 10^{-2}, 1$]~Hz, as~a
consequence of its smaller arm length. gLISA$_d$, on~the other hand,
matches the performance of gLISA only over the narrower frequency
region, [$7 \times 10^{-1}, 1$]~Hz, as~a consequence of its higher
acceleration noise at lower frequencies. The~angular, frequency, and
amplitude precisions as functions of the source sky location were then
derived by assuming an average signal-to-noise ratio of 10 at a
selected number of~{GW} frequencies covering the
operational bandwidth of TianQin and gLISA. Similar precision
functions were then derived for gLISA$_d$ by using the amplitudes
resulting in the gLISA average SNR of 10 at the same selected
frequencies. We found that, for~any given source location, all three
missions displayed a marked precision improvement in the three
reconstructed parameters at higher GW~frequencies.

{The precision levels presented in this article are based on the noise spectral densities presented in the literature for the TianQin and gLISA mission concepts. However, it should be noted that the key parameter that contains the noise and determines the precision values is the SNR. This can also be seen in the analytical expressions provided by \mbox{Equations~(\ref{DeltaOmega})--(\ref{Deltah}),} for~the precision of the parameters considered. An~increase or decrease in SNR will result in different parameter precisions, as~can be inferred from the aforementioned equations.}

Our analysis has shown that these three missions will be able to fill
the frequency gap between the region accessible by LISA and TaiJi and
that by ground interferometers. The~mid-band frequency region is
expected to contain a wide variety of sources of sinusoidal signals,
such as the white-dwarf--white-dwarf binary systems present in our
galaxy and hundreds of thousands to millions of binary black holes now
routinely observed by ground-based interferometers. The~GW signals
emitted by these systems can last for several months in the mid-band
frequency region accessible by these detectors, making them primary
candidates for detection and analysis, to~be then followed up with
ground-based interferometers. We plan to extend our analysis to
chirping signals emitted by these sources in a forthcoming~article.

\vspace{6pt} 

\authorcontributions{The authors of this article contributed equally
  to the work~reported. Conceptualization, M.F.S., T.A.F. and M.T.; methodology,  M.F.S., T.A.F. and M.T.; software,  M.F.S. and T.A.F.; validation,  M.F.S., T.A.F. and M.T.; formal analysis,  M.F.S., T.A.F. and M.T.; investigation,  M.F.S., T.A.F. and M.T.; resources,  M.F.S., T.A.F. and M.T.; data curation,  M.F.S., T.A.F. and M.T.; writing---original draft preparation,  M.F.S., T.A.F. and M.T.; writing---review and editing,  M.F.S., T.A.F. and M.T.; visualization, M.F.S. and T.A.F.; supervision, M.T.; project administration, M.T.; funding acquisition,  M.F.S., T.A.F. and M.T. All authors have read and agreed to the published version of the manuscript.}

\funding{For M.F.S., this research was funded by Conselho Nacional de
  Desenvolvimento Cient\'{i}fico e Tecnol\'{o}gico, grant
  No. 173535/2023-2. For~T.A.F., this research was funded by the
  National Science Foundation under grant number NSF-PHY2110509. For~  M.T., this research was funded by the Polish National Science Center
  Grant No.  2023/49/B/ST9/02777.}

\dataavailability{Data are contained within the article. 
}

\acknowledgments{We thank our institutions (UTFPR/UFPR, LSU and INPE) for their kind hospitality while this work was being carried out. }

\conflictsofinterest{The authors declare no conflicts of interest. The~  funders had no role in the design of the study; in the collection,
  analysis, or~interpretation of data; in the writing of the
  manuscript; or in the decision to publish the~results.}


\abbreviations{Abbreviations}{
The following abbreviations are used in this manuscript:\\
\noindent 
\begin{tabular}{@{}ll}
GW & Gravitational Wave\\
TDI & Time-Delay Interferometry
\end{tabular}
}

\vspace{6pt}

\appendixtitles{yes} 
\appendixstart
\appendix
\section[\appendixname~\thesection]{Michelson Response to a Sinusoidal Signal}
\label{append_A}

Based on the one-way Doppler time series given in
Equations~(\ref{eq:y21}) and (\ref{eq:y31}), we can obtain the two-way Doppler data
measured onboard spacecraft 1 and then derive the response of the
Michelson interferometer with equal-arm-length, $M_1$. The~following
expressions of the two-way Doppler responses measured onboard
spacecraft $1$ from arm $3$ ($Y_{21}$) and arm $2$ ($Y_{31}$) are
equal to
\begin{align} \label{eq:Y21}
Y_{21} &= y_{21}^{GW}(t) + y_{21,L}^{GW}(t) = y_{21}^{GW}(t) + y_{21}^{GW}(t - L)  \nonumber \\[5pt]
& = \left [ 1 - \frac{l}{L}(\mu_{1}-\mu_{2}) \right ] [\Psi_{3}(t-\mu_{1}l-2L)]  \\
& \qquad + \left [ 2\frac{l}{L_{3}}(\mu_{1}-\mu_{2}) \right ] [\Psi_{3}(t-\mu_{2}l-L)] \nonumber \\
& \qquad - \left [ 1 + \frac{l}{L}(\mu_{1}-\mu_{2}) \right ] [\Psi_{3}(t-\mu_{1}l)], \nonumber
\end{align}
and
\begin{align} \label{eq:Y31}
Y_{31} &= y_{31}^{GW}(t) + y_{31,L}^{GW}(t) = y_{31}^{GW}(t) + y_{31}^{GW}(t - L)  \nonumber \\[5pt]
& = \left [ 1 + \frac{l}{L}(\mu_{3}-\mu_{1}) \right ] [\Psi_{2}(t-\mu_{1}l-2L)]  \\
& \qquad - \left [ 2\frac{l}{L}(\mu_{3}-\mu_{1}) \right ] [\Psi_{2}(t-\mu_{3}l-L)] \nonumber \\
& \qquad - \left [ 1 - \frac{l}{L}(\mu_{3}-\mu_{1}) \right ] [\Psi_{2}(t-\mu_{1}l)] \ . \nonumber
\end{align}

From the above two-way Doppler combinations, it is then possible to obtain the following expression for the equal-arm-length Michelson combination $M_1$:

\begin{align} \label{eq:M1}
M_{1}(t) &= Y_{21} - Y_{31}  \nonumber \\[5pt]
& = \left [ 1 + \frac{l}{L}(\mu_{2}-\mu_{1}) \right ] [\Psi_{3}(t-\mu_{1}l-2L)] - \left [ 1 + \frac{l}{L}(\mu_{3}-\mu_{1}) \right ] [\Psi_{2}(t-\mu_{1}l-2L)]  \nonumber \\
& \quad - \left [ 2\frac{l}{L}(\mu_{2}-\mu_{1}) \right ] [\Psi_{3}(t-\mu_{2}l-L)] + \left [ 2\frac{l}{L}(\mu_{3}-\mu_{1}) \right ] [\Psi_{2}(t-\mu_{3}l-L)] \nonumber \\
& \quad - \left [ 1 - \frac{l}{L}(\mu_{2}-\mu_{1}) \right ] [\Psi_{3}(t-\mu_{1}l)] + \left [ 1 - \frac{l}{L}(\mu_{3}-\mu_{1}) \right ] [\Psi_{2}(t-\mu_{1}l)]. 
\end{align}

Since $L \, \hat{k} \cdot \hat{n}_{2} = l(\mu_{3}-\mu_{1})$, and~so forth by cyclic permutation of the indices, $M_{1}(t)$ can be rewritten as follows:
\begin{align} \label{eq:M1_kn}
M_{1}(t) &= \left ( 1 - \hat{k} \cdot \hat{n}_{3} \right ) [\Psi_{3}(t-\mu_{1}l-2L)] - \left ( 1 + \hat{k} \cdot \hat{n}_{2} \right ) [\Psi_{2}(t-\mu_{1}l-2L)] \nonumber \\
& \quad + \left ( 2\hat{k} \cdot \hat{n}_{3} \right ) [\Psi_{3}(t-\mu_{2}l-L)] + \left ( 2\hat{k} \cdot \hat{n}_{2} \right ) [\Psi_{2}(t-\mu_{3}l-L)]  \nonumber \\
& \quad - \left ( 1 + \hat{k} \cdot \hat{n}_{3} \right ) [\Psi_{3}(t-\mu_{1}l)] + \left ( 1 - \hat{k} \cdot \hat{n}_{2} \right ) [\Psi_{2}(t-\mu_{1}l)].
\end{align}

The terms $\Psi_{i}(t)$ can be further expanded to explicitly show their dependence on the parameters characterizing the~{GW} signal. Thus, substituting ${\sf H}(t)$ into the expression of
$\Psi_{i}(t)$, we obtain
\begin{equation}
\Psi_{i}(t) =  \frac{\hat{n}_{i} \cdot [h_{+} \mathbf{e_{+}} + h_{\times} \mathbf{e_{\times }}] \cdot \hat{n}_{i}}{2[1-(\hat{k} \cdot \hat{n}_{i})^{2}]}
= \frac{\hat{n}_{i} \cdot [h_{0+} \cos{(\omega t)} \, \mathbf{e_{+}} + h_{0 \times} \sin{(\omega t)} \, \mathbf{e_{\times }}] \cdot \hat{n}_{i}}{2[1-(\hat{k} \cdot \hat{n}_{i})^{2}]},
\label{eq:PSI_exp0}
\end{equation}
\begin{equation}
\Psi_{i}(t) =  h_{0+} \cos{(\omega t)} \, \Psi_{i+} + h_{0 \times} \sin{(\omega t)} \, \Psi_{i \times},
\label{eq:PSI_exp1}
\end{equation}
where $\Psi_{i+}$ and $\Psi_{i \times}$ are given by
\begin{equation}
\Psi_{i+} =  \frac{\hat{n}_{i} \cdot \mathbf{e_{+}} \cdot \hat{n}_{i}}{2[1-(\hat{k} \cdot \hat{n}_{i})^{2}]}
\hspace{1.5cm}
\Psi_{i\times} =  \frac{\hat{n}_{i} \cdot \mathbf{e_{\times }} \cdot \hat{n}_{i}}{2[1-(\hat{k} \cdot \hat{n}_{i})^{2}]}.
\label{eq:Psi_pc}
\end{equation}

Note the expression of the equal-arm-length Michelson response,
Equation~(\ref{eq:M1_kn}), displays the characteristic ``four-pulse''
structure~\cite{AET99}, as~the~{GW} signal appears in it
at four distinct delay times. Let us denote these time delays as
$\tau_{a}$, where the subscript $a$ represents the various
combinations of these retarded times. Thus,
Equation~(\ref{eq:PSI_exp1}) can be written as
\begin{align} \label{eq:PSI_tau}
\Psi_{i}(t - \tau_{a}) &=  h_{0+} \Psi_{i+} \cos{(\omega t - \omega \tau_{a})} + h_{0 \times} \Psi_{i \times} \sin{(\omega t- \omega \tau_{a})}   \nonumber \\[5pt]
& = h_{0+} \cos{(\omega t)} \left [ \Psi_{i+} \cos{(\omega \tau_{a})} - \mathcal{A} \Psi_{i \times} \sin{(\omega \tau_{a})} \right ]  \nonumber \\
& \quad + h_{0+} \sin{(\omega t)} \left [ \Psi_{i+} \sin{(\omega \tau_{a})} + \mathcal{A} \Psi_{i \times} \cos{(\omega \tau_{a})} \right ], 
\end{align}
where we have defined $\mathcal{A} \equiv h_{0 \times} / h_{0+}$.  By~substituting the $\Psi_{i}(t - \tau_{a})$ terms from
\mbox{Equation~(\ref{eq:PSI_tau})} into Equation~(\ref{eq:M1_kn}) of the Michelson
interferometer, and~after performing some algebraic manipulations, we
obtain the response of a stationary Michelson interferometer to a
sinusoidal GW signal:
\begin{equation}
M_{1}(t) = h_{0+} \cos{(\omega t)} \, F_{\rm I} + h_{0+} \sin{(\omega t)} \, F_{\rm II},
\label{eq:M1_gnral01}
\end{equation}
where 
\begin{align} \label{eq:F_plus01}
F_{\rm I} &= \left ( 1 - \hat{k} \cdot \hat{n}_{3} \right ) [\Psi_{3+} \cos{(\omega \tau_{1})} - \mathcal{A} \Psi_{3 \times} \sin{(\omega \tau_{1})}] \nonumber \\
& \quad - \left ( 1 + \hat{k} \cdot \hat{n}_{2} \right ) [\Psi_{2+} \cos{(\omega \tau_{1})} - \mathcal{A} \Psi_{2 \times} \sin{(\omega \tau_{1})}] \nonumber \\
& \quad + \left ( 2\hat{k} \cdot \hat{n}_{3} \right ) [\Psi_{3+} \cos{(\omega \tau_{2})} - \mathcal{A} \Psi_{3 \times} \sin{(\omega \tau_{2})}] \nonumber \\
& \quad + \left ( 2\hat{k} \cdot \hat{n}_{2} \right ) [\Psi_{2+} \cos{(\omega \tau_{3})} - \mathcal{A} \Psi_{2 \times} \sin{(\omega \tau_{3})}] \nonumber \\
& \quad - \left ( 1 + \hat{k} \cdot \hat{n}_{3} \right ) [\Psi_{3+} \cos{(\omega \tau_{4})} - \mathcal{A} \Psi_{3 \times} \sin{(\omega \tau_{4})}] \nonumber \\
& \quad + \left ( 1 - \hat{k} \cdot \hat{n}_{2} \right ) [\Psi_{2+} \cos{(\omega \tau_{4})} - \mathcal{A} \Psi_{2 \times} \sin{(\omega \tau_{4})}], 
\end{align}
and
\begin{align} \label{eq:F_cross01}
F_{\rm II} &= \left ( 1 - \hat{k} \cdot \hat{n}_{3} \right ) [\Psi_{3+} \sin{(\omega \tau_{1})} + \mathcal{A} \Psi_{3 \times} \cos{(\omega \tau_{1})}] \nonumber \\
& \quad - \left ( 1 + \hat{k} \cdot \hat{n}_{2} \right ) [\Psi_{2+} \sin{(\omega \tau_{1})} + \mathcal{A} \Psi_{2 \times} \cos{(\omega \tau_{1})}]  \nonumber \\
& \quad + \left ( 2\hat{k} \cdot \hat{n}_{3} \right ) [\Psi_{3+} \sin{(\omega \tau_{2})} + \mathcal{A} \Psi_{3 \times} \cos{(\omega \tau_{2})}] \nonumber \\
& \quad + \left ( 2\hat{k} \cdot \hat{n}_{2} \right ) [\Psi_{2+} \sin{(\omega \tau_{3})} + \mathcal{A} \Psi_{2 \times} \cos{(\omega \tau_{3})}]  \nonumber \\
& \quad - \left ( 1 + \hat{k} \cdot \hat{n}_{3} \right ) [\Psi_{3+} \sin{(\omega \tau_{4})} + \mathcal{A} \Psi_{3 \times} \cos{(\omega \tau_{4})}] \nonumber \\
& \quad + \left ( 1 - \hat{k} \cdot \hat{n}_{2} \right ) [\Psi_{2+} \sin{(\omega \tau_{4})} + \mathcal{A} \Psi_{2 \times} \cos{(\omega \tau_{4})}], 
\end{align}
where the time delays ($\tau_{a} \ , \ a = 1, 2, 3, 4$) are equal to
$\tau_{1} = (\mu_{1}l+2L)$, $\tau_{2} = (\mu_{2}l+L)$,
\mbox{$\tau_{3} = (\mu_{3}l+L)$ and $\tau_{4} = \mu_{1}l$. } Following a
similar procedure as described above, the~responses of the other two
Michelson interferometers, $M_{2}(t)$ and $M_{3}(t)$ measured onboard
spacecraft $2$ and $3$, respectively, can be~derived.

\vspace{8pt} 



\begin{adjustwidth}{-\extralength}{0cm}

\printendnotes[custom]

\reftitle{References}


\PublishersNote{}
\end{adjustwidth}
\end{document}